\documentclass[a4paper,twocolumn,11pt,accepted=2026-03-02]{quantumarticle}
\pdfoutput=1
\usepackage[utf8]{inputenc}
\usepackage[english]{babel}
\usepackage[T1]{fontenc}

\usepackage{lipsum}
\usepackage{mdframed}
\usepackage{booktabs}
\usepackage{longtable}

\usepackage[colorlinks]{hyperref}
\usepackage[page,header]{appendix}

\usepackage{amsmath,amsfonts,amssymb,amstext, amsthm}
\usepackage[sort&compress, numbers]{natbib}
\usepackage[pdftex]{graphicx}
\usepackage{xcolor}
\usepackage{comment}
\usepackage{bbm}

\usepackage{mathtools}
\usepackage{nccmath}
\usepackage{blkarray}
\usepackage{here}
\usepackage{array}
\usepackage{tikz}

\newtheorem{theorem}{Theorem}
\newtheorem{proposition}[theorem]{Proposition}
\newtheorem{lemma}[theorem]{Lemma}
\newtheorem{corollary}[theorem]{Corollary}

\theoremstyle{definition}

\theoremstyle{definition}

\usepackage[normalem]{ulem}

\newcommand{\ket}[1]{|#1\rangle} 
\newcommand{\bra}[1]{\langle#1|} 
\newcommand{\ketbra}[2]{|#1\rangle\langle#2|} 
\newcommand{\braket}[2]{\langle #1 \vert #2 \rangle}

\DeclareMathOperator{\tr}{Tr}

\newcommand{\cC}{{\mathcal{C}}}
\newcommand{\cD}{{\mathcal{D}}}
\newcommand{\cE}{{\mathcal{E}}}
\newcommand{\cF}{{\mathcal{F}}}
\newcommand{\cG}{{\mathcal{G}}}
\newcommand{\cH}{{\mathcal{H}}}

\newcommand{\cM}{{\mathcal{M}}}
\newcommand{\cN}{{\mathcal{N}}}
\newcommand{\cO}{{\mathcal{O}}}
\newcommand{\cP}{{\mathcal{P}}}

\newcommand{\cR}{{\mathcal{R}}}
\newcommand{\cS}{{\mathcal{S}}}
\newcommand{\cT}{{\mathcal{T}}}

\newcommand{\cV}{{\mathcal{V}}}

\newcommand{\sT}{\mathsf{T}}
\newcommand{\bI}{\mathbb{I}}

\newcommand{\f}{\frac}

\newcommand{\rarr}{\rightarrow}

\newcommand{\til}{\tilde}

\newcommand{\sign}{\mathrm{sign}}

\begin{document}

\begin{flushright}
YITP-24-60 
\end{flushright}

\title{Explicit decoders using fixed-point amplitude amplification based on QSVT}

\author{Takeru Utsumi}
\email{takeru-utsumi@g.ecc.u-tokyo.ac.jp}
\affiliation{Graduate School of Arts and Sciences, The University of Tokyo, Komaba, Meguro-ku, Tokyo 153-8902, Japan}
\orcid{0009-0004-2935-9794}

\author{Yoshifumi Nakata}
\email{yoshifumi.nakata@yukawa.kyoto-u.ac.jp}
\affiliation{Yukawa Institute for Theoretical Physics, Kyoto University, Kitashirakawa, Sakyo-ku, Kyoto 606-8502, Japan}
\orcid{0000-0003-1285-6968}

\maketitle


\begin{abstract}
Reliably transmitting quantum information via a noisy quantum channel is a central challenge in quantum information science. While constructing a decoder is crucial to this goal, little was known about quantum circuit implementations of decoders that reach high communication rates. In this paper, we provide two decoders with explicit quantum circuits capable of recovering quantum information when the decoupling condition is satisfied, i.e., when quantum information is in principle recoverable. These are applicable to both entanglement-assisted and non-assisted settings. By developing a technique that relies on a symmetric structure of the decoders, we show that they are applicable to any noise model. As a consequence, for any noisy channel, our decoders can be used to achieve a communication rate arbitrarily close to the quantum capacity by increasing the number of channel uses. To construct the decoders, we employ the fixed-point amplitude amplification (FPAA) based on the quantum singular value transformation (QSVT), extending a previous approach applicable only to erasure noise. Our constructions offer advantages in the computational cost, largely reducing the circuit complexity compared to previous explicit decoders. Through an investigation of the decoding problem, unique advantages of the QSVT-based FPAA are highlighted.
\end{abstract}

\section{Introduction}
\label{sec: introduction}

Protecting quantum information from the effects of noise is crucial for transmitting quantum information over noisy quantum channels.
A standard technique is to use quantum error correcting codes (QECCs), in which quantum information is encoded before the system experiences noise and is decoded afterward. 
The QECCs are also of significant importance in fundamental physics, such as the black hole information paradox~\cite{hayden2007black, harlow2013CompvsFirewall, nakata2023black}, the AdS/CFT correspondence~\cite{pastawski2015holographic, almheiri2015bulk}, topological orders~\cite{dennis2002topological, kitaev2003fault, kitaev2006anyons}, and quantum chaos~\cite{hosur2016chaos,roberts2017chaos, nakata2024haydenpreskill}.

The importance of high-performance QECCs, especially those capable of transmitting quantum information at high rates, is growing with the recent advancement of quantum technology.
In particular, achieving the optimal communication rate, i.e., the \emph{quantum capacity}, is one of the main goals of quantum information theory.
A standard approach to this goal relies on the concept of \emph{decoupling}~\cite{hayden2008decoupling, dupuis2010decoupling, dupuis2014one}, which provides a necessary and sufficient condition for the recoverability of quantum information from a noisy system. 
The decoupling approach is, however, for investigating the capability of encoders and offers little insight into explicit decoding strategies. 
Explicitly constructing decoders whose rate approaches the quantum capacity has long been a central challenge.

Some progress has been made so far on QECCs with specific structures aimed at achieving high communication rates. For instance, a Clifford-based polar code provides an explicit decoder for Pauli noise that reaches a rate near the quantum capacity~\cite{dupuis2021polarQCclifford}. 
In another approach, \cite{Renes2017BPpassingmessages} and subsequent works~\cite{Rengaswamy2021BRquantclassical, Piveteau2022quantummessage} developed belief-propagation decoders for pure-state classical-quantum channels, which achieve the classical capacity.
There is also an approach based on CSS-type encoding with decoding via measurements of complementary observables~\cite{Renes2013thephysics, Renes2022QuantumInformationCM}.
Despite the significance of the decoding problem, however, only a handful of results have been obtained in general cases~\cite{barnum2002reversing, renes2016uncertain, nakata2025constructing} due to its inherent difficulties.
Exploring general decoders can be beneficial towards achieving high-performance QECCs or the quantum capacity.

Among the previously proposed decoders applicable to general noise, to the best of our knowledge, the only one known to be explicit and approximately achieve the quantum capacity is based on the Petz recovery map~\cite{petz1986sufficient, petz1988sufficiency}.
Since the map is known to attain near-optimal recovery error~\cite{barnum2002reversing}, and also to exhibit favorable second-order performance~\cite{beigi2016decoding}, it can serve as a high-performance decoder.
However, implementing the Petz recovery map with a quantum circuit requires significant computational costs~\cite{yoshida2021decoding, gilyen2022petzmap, biswas2023noiseadapted, nakayama2023petz}. From a practical point of view, it is desirable to develop explicit quantum circuit constructions with reduced computational costs.

An explicit decoder with smaller computational cost is possibly constructed by following the approach of Yoshida and Kitaev~\cite{yoshida2017efficient}.
They proposed a decoding protocol, which we call the \emph{Yoshida--Kitaev (YK) decoder}, for the Hayden--Preskill (HP) protocol~\cite{hayden2007black}, a specific noisy model of the qubit-erasure noise with a unitary encoding, relevant to a toy model of a quantum black hole.
The proposal consists of two steps: first, a decoding protocol with post-selection is considered, and then the protocol is lifted up to a decoding quantum circuit without post-selection through a non-trivial use of the \emph{amplitude amplification (AA)} algorithm~\cite{grover1996stoc, Brassard1997Simons, brassard2002quantum}.
This two-step construction provides an explicit quantum circuit and successfully decodes the HP protocol, contributing to high-energy physics (see, e.g., \cite{Li2024retrevalBH, Li2024quantuminffromBH}). However, the second step with the AA algorithm, as well as the proof technique, is specifically tailored to the HP protocol. As a result, the approach is not applicable to other noise models, limiting its impact on quantum information theory.
This naturally raises the question: \emph{can the two-step construction be extended to construct a decoder applicable to an arbitrary noise model?}
If this question is answered affirmatively and the resulting decoder achieves a high communication rate, it contributes to the field of quantum information.

The key to extending the construction would lie in upgrading the step with the AA using recently proposed quantum algorithms. 
In the past decade, significant progress has been made in quantum algorithms, including the AA algorithm~\cite{grover1996stoc, Brassard1997Simons, brassard2002quantum}. The \emph{fixed-point amplitude amplification (FPAA)}~\cite{grover2005fixed, aaronson2012mony, yoder2014fixed} is one of the improved ones applicable to broader situations.
An approach to the FPAA is based on the \emph{quantum singular value transformation (QSVT)}~\cite{gilyen2019qsvt, Gilyn2019thesis, martyn2021grand}, exhibiting a unique property not found in other AA-type algorithms (see~\ref{sec: explanation QSVT-based FPAA}). 
The distinctive feature of the QSVT-based FPAA may help achieve tasks beyond the reach of any other AA-type algorithms.
This observation may also be of theoretical interest as it will offer concrete examples in applications that highlight the unique advantages of the QSVT-based FPAA.

In this paper, by building upon these recent advances, we construct two explicit decoding circuits applicable to arbitrary noisy channels.
One is a \emph{generalized YK} decoder, and the other is a \emph{Petz-like} decoder. Both of them are obtained by upgrading the two-step construction with the unique advantages of the QSVT-based FPAA.
The Petz-like decoder is a more concise alternative to the Petz recovery map, suggesting that a full implementation of the map is not necessary for decoding.
These decoders are applicable to both the entanglement-assisted and non-assisted settings.
We show that these decoders can recover quantum information when the decoupling condition is satisfied. 
Since the decoupling is necessary and sufficient for recovery, the decoders succeed in recovering quantum information whenever it is in principle possible.
This implies that, in the i.i.d. setting, both decoders with suitably chosen encoders asymptotically achieve a rate arbitrarily close to the quantum capacity.
The technical contribution lies in a novel proof method relying solely on a general structure of the construction, by which the original method's limitations in applicability can be circumvented. Specifically, we exploit a symmetry of the two-step construction and apply the Powers--St\o rmer inequality~\cite{powers1970free, kittaneh1987inequalities}, a powerful tool of mathematical physics.

Furthermore, taking advantage of explicit constructions, we provide an in-depth analysis of the circuit complexity of the decoders. We show that both of our decoding circuits largely reduce the circuit complexity.
The complexity remains exponential in the number of qubits, which should be the case due to the inherent computational hardness of a general decoding problem~\cite{Vardy1997intractability, Kuo2012hardnessdepo, Iyer2015hardness}. 
Nevertheless, we have succeeded in reducing the exponent of the exponential scaling compared to the previously known explicit decoding circuits~\cite{gilyen2022petzmap}, resulting in a significant improvement. 
We also provide a simple criterion to determine which of the two decoders, the generalized YK and the Petz-like decoders, has smaller complexity.
Although the criterion depends on many factors, we show that the generalized YK decoder always has smaller complexity when an encoder is an isometry and noisy channels have the maximum number of Kraus operators.

Our constructions, among the first few to provide decoders achieving a rate arbitrarily close to the quantum capacity with explicit quantum circuit implementations, contribute to the main challenge in quantum information theory. Additionally, from a quantum algorithms perspective, our construction works with the QSVT-based FPAA but fails with other known AA-type algorithms, clearly exemplifying the separation between these algorithms in a concrete and practically relevant problem. 
Although such a separation was previously pointed out~\cite{Gilyn2019thesis}, to our knowledge, few concrete examples were found in the literature.
Moreover, our example highlights the situation in which the separation becomes particularly pronounced. Specifically, the unique advantages of the QSVT-based FPAA stand out when the algorithm is necessarily applied to a subsystem of an entangled system, which is precisely the case in the decoding problem. This observation provides insights into future applications of the QSVT-based algorithms.

This paper is organized as follows. We start with preliminaries in \ref{sec: preliminaries}. Our main results are summarized in \ref{sec: main results}. The proofs of our results are provided in \ref{sec: proofs of these results}. We conclude with a summary and outlooks in \ref{sec: conclusion}.

\section{Preliminaries}
\label{sec: preliminaries}

We here introduce our notation and our setting. We then briefly overview an implicit decoder commonly used in the decoupling approach. We also provide quick overviews of the Petz recovery map and the original two-step construction of the YK decoder for the HP protocol~\cite{yoshida2017efficient}.
We then concisely highlight the properties of several known AA-type algorithms and provide a brief introduction to the QSVT-based FPAA.

\subsection{Notation}

\begin{figure*}
    \centering
    \includegraphics[width=160mm]{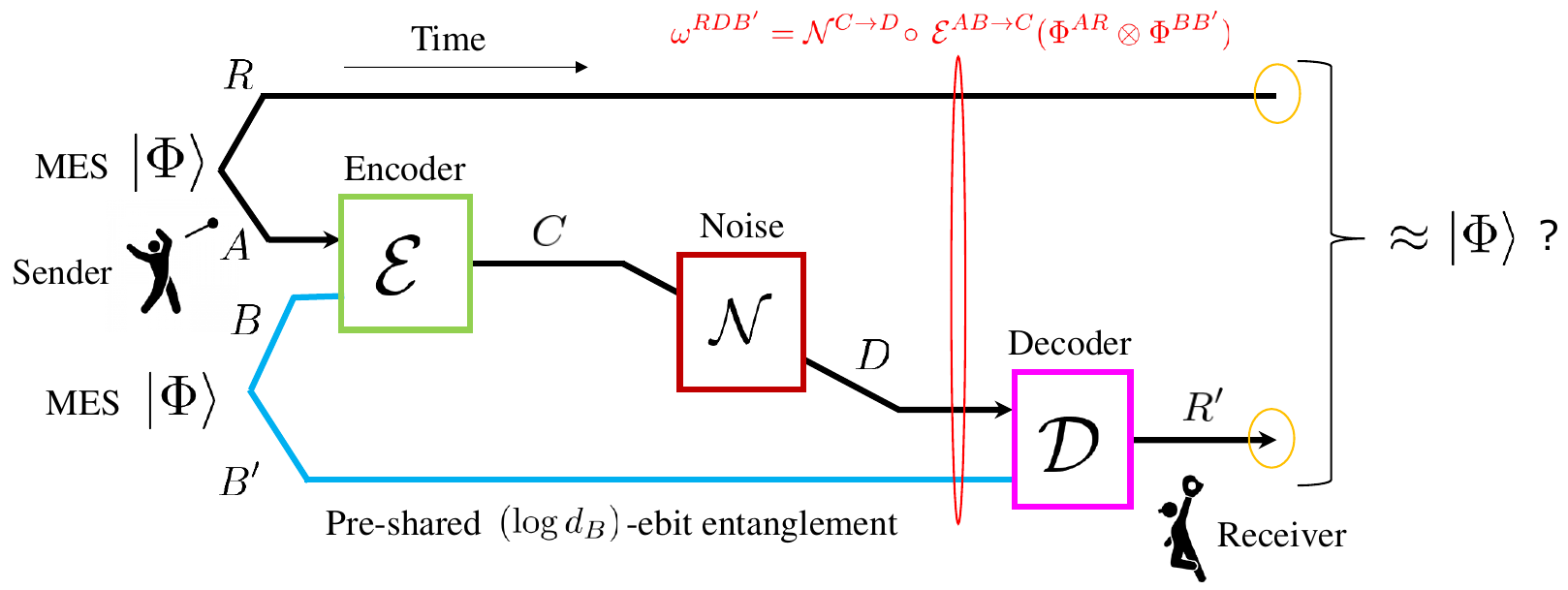}
    \vspace{-2mm}\caption{A diagram of our setting. Time flows from left to right. The boxes represent quantum channels. The purpose of the sender and the receiver is to transmit quantum information via a noisy channel $\cN^{C\mapsto D}$, which is equivalent to preserving the maximally entangled state between $A$ and $R$. They may share $(\log d_B)$-ebit entanglement in advance, which is used during the encoding and decoding operations. When $d_B = 1$, this corresponds to the entanglement-non-assisted setting, while $d_B \neq 1$ corresponds to the entanglement-assisted setting with a limited or unlimited amount of entanglement, depending on whether $d_B$ is bounded or arbitrarily large, respectively.}
\label{fig: ent asist}
\end{figure*}

Throughout this paper, we denote by $\mathcal{S}(\mathcal{H})$ a set of all quantum states on a Hilbert space $\mathcal{H}$. While we usually denote a pure state by $\ket{\varphi}$, the corresponding density operator is sometimes described as $\varphi$, namely, $\varphi = \ketbra{\varphi}{\varphi}$. We use a superscript to represent a system on which operators and maps are defined. For instance, an operator on a system $AB$ and a superoperator from $A$ to $B$ are denoted by $\varphi^{AB}$ and $\mathcal{T}^{A \rarr B}$, respectively. The superscript is omitted when it is clear from the context. A reduced density operator on $A$ of $\varphi^{AB}$ is described as $\varphi^A$, i.e., $\varphi^A = \tr_B \varphi^{AB}$, where $\tr_B$ is the partial trace over $B$. 

For an operator $M$, we denote the complex conjugate and the transpose in a given basis by $M^*$ and $M^\mathsf{T}$, respectively, and denote the Hermitian conjugate by $M^\dag$. The identity operation is denoted by $\mathbb{I}$ and ${\rm id}$ for operators and superoperators, respectively. We often omit the identity operators and superoperators for simplicity. 

A Hilbert space, such as $\cH^{A'}$ or $\cH^{\hat{A}}$, is isomorphic to $\cH^A$: it has the same dimension and the same fixed basis as $\cH^A$. This applies not only to the system $A$, but also to any systems, such as $\cH^{B'}$ and $\cH^{\hat{C}}$. We write the dimension of a Hilbert space $\cH$ as $d$, and for instance, denote by $d_A$ the dimension of $\cH^A$. 

We omit the symbol of the tensor product between vectors and denote it as $\ket{\varphi}\otimes\ket{\psi} = \ket{\varphi}\ket{\psi}$, for simplicity, when it is clear from the context.
We denote by $\ket{\Phi}$ a maximally entangled state (MES) defined in an orthonormal computational basis. For instance, the MES between $A$ and $\hat{A}$ is 
\begin{equation}
    \ket{\Phi}^{A\hat{A}} = \f{1}{\sqrt{d_A}}\sum_{i=1}^{d_A}\ket{i}^A\ket{i}^{\hat{A}},
\end{equation}
where $\{\ket{i}\}_i$ is the computational basis in $A$ and $\hat{A}$, respectively. 
Note that a MES in an arbitrary basis can be transformed into the MES in the computational basis by applying an appropriate unitary to one of the local systems. 
Also, note that $\sqrt{d_A}\ket{\Phi}^{AA'} = \sum_i \ket{i}^A\ket{i}^{A'} = \sum_i\ket{e_i}^A\ket{e_i^*}^{A'}$ holds for any orthonormal basis $\{\ket{e_i}^A\}_i$. Here, the complex conjugate is taken with respect to the computational basis; that is, when $\ket{e_i}^A = \sum_j c_{ij} \ket{j}^A$, it is given by $\ket{e_i^*}^{A'} = \sum_j c_{ij}^* \ket{j}^{A'}$.
We denote the completely mixed state (CMS) by $\pi$, such as $\pi^A = \mathbb{I}^A/d_A$.

The circuit complexity of $\mathcal{T}$ is denoted by $\mathcal{C}(\mathcal{T})$. This is the minimum total number of single- and two-qubit unitary gates required to perform $\mathcal{T}$, allowing for ancillae polynomial in the number of qubits.

For a matrix $M$, the Schatten-$p$ norm is defined by $\| M \|_p \coloneqq \big(\tr\big[(\sqrt{M^\dag M})^p\big]\big)^{1/p}$. 
We particularly use the trace norm, the Hilbert--Schmidt norm, and the operator norm, corresponding to $p = 1, 2, \infty$, respectively.
The trace norm has the contraction property such that for $\varphi^{AB} \in \cS(\cH^{AB})$ and $\psi^{AB} \in \cS(\cH^{AB})$,
\begin{equation}
\label{eq: contraction trace norm}
    \|\varphi^A - \psi^A\|_1 \leq \|\varphi^{AB} - \psi^{AB}\|_1.
\end{equation}
The fidelity between $\varphi \in \cS(\cH)$ and $\psi \in \cS(\cH)$ is defined by $\mathrm{F}(\varphi, \psi) \coloneqq \big\|\sqrt{\varphi}\sqrt{\psi}\big\|_1^2$. 
The fidelity is rephrased using the purified states of $\varphi$ and $\psi$ as 
\begin{equation}
\label{eq: Uhlmann thm}
    \mathrm{F}(\varphi^A, \psi^A) = \max_{V}\big| \bra{\varphi}^{AC}V^{B\rarr C}\ket{\psi}^{AB} \big|^2,
\end{equation}
where the maximization is taken over all isometries $V^{B\rarr C}$. We supposed $d_C \geq d_B$ without loss of generality. This is called the \emph{Uhlmann's theorem} \cite{uhlmann1976transition}.
The trace norm and the fidelity are related by the Fuchs--van de Graaf inequalities \cite{Fuchs1999fuchsvandegraaf, watrous2018TheoryQI}: 
\begin{equation}
\label{eq: fuchs van de graaf}
   1-\sqrt{\mathrm{F}(\varphi, \psi)} \leq \f{1}{2}\|\varphi - \psi \|_1 \leq \sqrt{1-\mathrm{F}(\varphi, \psi)}. 
\end{equation}

We use the quantum collision entropy. For $\varphi^A \in \cS(\cH^A)$ it is given by 
\begin{equation}
    H_2(A)_\varphi = - \log\tr[(\varphi^A)^2].
\end{equation}
This satisfies $0 \leq H_2(A)_\varphi \leq \log d_A$.


\subsection{Our setting}

We consider the following general setting, which is common in quantum communication. See also Fig.~\ref{fig: ent asist}.
A sender aims to transmit ($\log d_A$)-qubit quantum information using a given noisy channel $\cN^{C\rarr D}$ and possibly a pre-shared entanglement $\ket{\Phi}^{BB'}$, where $B$ and $B'$ are with the sender and receiver, respectively. 
When they share no entanglement, we set $d_B = 1$ and call this the entanglement-non-assisted setting. Otherwise, it is the entanglement-assisted setting, with a limited or unlimited amount of entanglement depending on whether $d_B$ is bounded or arbitrarily large, respectively.
The sender encodes the system $A$ with $B$ using an encoding channel $\cE^{AB\rarr C}$. The qubits in $C$ are then transmitted to the receiver through the noisy channel $\cN^{C\rarr D}$. 
The receiver obtains the output system $D$ of the noisy channel and applies a recovery channel, i.e., a decoder $\cD^{DB'\rarr R'}$ onto the system $DB'$.
For simplicity, we denote by $\cF^{AB \rarr D}$ the composite channel $\cN^{C\rarr D} \circ \cE^{AB \rarr C}$.

Following the convention, we introduce a reference system $R$ isomorphic to $A$ with $d_R =d_A$, and prepare the systems $A$ and $R$ to be in a MES $\ket{\Phi}^{AR}$.
We denote by $\omega^{RDB'}$ the state after the noise, i.e., just before the decoder: 
\begin{equation}
\label{eq: pre noisy state}
    \omega^{RDB'} \coloneqq \cF^{AB \rarr D}(\Phi^{AR}\otimes \Phi^{BB'}).
\end{equation}
The recovery error of quantum information by a decoder $\cD^{DB' \rightarrow R'}$ is defined as~\footnote{It is known that the recovery errors defined by other metrics, such as the diamond norm, are closely related to the recovery error we adopt~\cite{kretschmann2004tema, devetak2005private, watrous2018TheoryQI, nakata2025constructing, khatri2024principlemodern}.}
\begin{equation}\begin{split}
\label{eq: def. recovery error}
    \Delta(\cD|\cF)  \coloneqq \f{1}{2}\|\Phi^{RR'} - \cD^{DB' \rarr R'}(\omega^{RDB'}) \|_1.
\end{split}\end{equation}

The main concern in this paper is to explicitly construct a decoder $\cD^{DB'\rarr R'}$ for a given channel $\cF^{AB\rarr D}$.
We assume that the descriptions of the encoding map $\cE$ and the noisy channel $\cN$ are known, so that the decoder can depend on their details.
It is also assumed that every operation, except for the noisy channel $\cN$, can be performed noiselessly. This is a common assumption in studies of information transmission. For practical implementation, however, it is also important to relax this assumption, which we will mention in~\ref{sec: conclusion}.


\subsection{Decoupling and quantum capacity}
\label{sec: decoupling and Uhlmann}

A standard approach to evaluating the recovery error is to estimate how much quantum information is leaked to an ``environment'' of the noisy channel. This is specifically quantified by the degree of decoupling.

Let $V_\cF^{AB \rarr ED}$ be a Stinespring isometry~\cite{stinespring1955} of the channel $\cF^{AB \rarr D} = \cN^{C\rarr D}\circ\cE^{AB \rarr C}$ by an environment $E$.
That is, the channel $\cF^{AB\rarr D}$ is represented as
\begin{equation}\begin{split}
\label{eq: stine dilation by iso}
    \cF^{AB\rarr D}(\cdot) 
    &= \tr_E\big[V_\cF^{AB\rarr ED}(\cdot)(V_\cF^{AB\rarr ED})^\dag\big].
\end{split}\end{equation}
For convenience, we also introduce a purified state of $\omega^{RDB'}$ in Eq.~\eqref{eq: pre noisy state} as 
\begin{equation}\begin{split}
        \ket{\omega}^{REDB'} &\coloneqq V_\cF^{AB \rarr ED}\ket{\Phi}^{AR} \ket{\Phi}^{BB'}. 
\end{split}\end{equation}
The following is called the decoupling approach.
\begin{proposition}[Decoupling approach \cite{hayden2008decoupling, dupuis2010decoupling, dupuis2014one}]
\label{prop: uhlmann's dec}
    Suppose $\ket{\omega}^{REDB'}$ is a pure state. If there exists a state $\tau^E$ such that $\|\omega^{RE} - \pi^R \otimes \tau^E\|_1 \leq \epsilon$, then there exists a quantum channel $\cD_{\rm Uhlmann}^{DB'\rarr R'}$ that satisfies 
    \begin{equation}
        \f{1}{2}\big\|\Phi^{RR'} - \cD_{\rm Uhlmann}^{DB'\rarr R'}(\omega^{RDB'}) \big\|_1 \leq \sqrt{\epsilon}.
    \end{equation}
\end{proposition}

This proposition is a straightforward consequence of the Uhlmann's theorem (Eq.~\eqref{eq: Uhlmann thm}) together with Eqs.~\eqref{eq: contraction trace norm} and~\eqref{eq: fuchs van de graaf}. See, e.g., \cite{hayden2008decoupling, dupuis2010decoupling, dupuis2014one}. 
The condition that there exists $\tau^{E}$ such that 
\begin{equation}
\label{eq: decp cond}
    \|\omega^{RE} - \pi^R \otimes \tau^{E}\|_1 \leq \epsilon,
\end{equation}
is called a \emph{decoupling condition}, and, in fact, it is known to be necessary and sufficient for the recoverability of quantum information.

The decoupling approach is particularly strong in the study of the maximum possible communication rate of quantum information, i.e., the quantum capacity, either in one-shot or asymptotic settings.
We now briefly describe the relation between the decoupling condition and the quantum capacity, as well as the construction of an explicit decoder.

Suppose that quantum information is transmitted through $N$ independent and identical uses of a noisy channel $\cN_1^{A_1 \rarr C_1}$, namely, $\cN^{A \rarr C} = (\cN_1^{A_1 \rarr C_1})^{\otimes N}$.
An (asymptotically) achievable rate is given by $\mathrm{R} \coloneqq \lim_{N\rarr \infty}\f{1}{N}\log d_A$ under the assumption that there exists a sequence of pairs of an encoder and a decoder such that the recovery error tends to zero as $N \rarr \infty$.
The quantum capacity is defined as the supremum of the achievable rate for the channel $\cN_1$.

As established in the quantum capacity theorem~\cite{lloyd1997capacity, barnum1998inftrans, Barnum2000OnquantumFidel, shor2002quantum, devetak2005private}, when the sender and receiver share no entanglement in advance, the quantum capacity $Q(\cN_1)$ is given by the regularized coherent information:
\begin{align}
    Q(\cN_1) = \lim_{N\rarr\infty}\f{1}{N}I_{\rm c}\big((\cN_1^{A_1 \rarr C_1})^{\otimes N}\big),
\end{align}
where $I_{\rm c}(\cT^{A \rarr B}) = \max_{\rho^A} \big[H\big(\cT^{A \rarr B}(\rho^A)\big) - H\big(\bar{\cT}^{A \rarr E}(\rho^A)\big)\big]$ is the coherent information of a quantum channel $\cT^{A \rarr B}$~\cite{wilde2013QItheory, watrous2018TheoryQI, khatri2024principlemodern}.
Here, $H(\rho^A) = -\tr[\rho^A \log \rho^A]$ is the von Neumann entropy of a state $\rho^A$ and $\bar{\cT}^{A\rarr E}$ is a complementary channel of $\cT^{A\rarr B}$.
The maximization is taken over all states on the input system of $\cT^{A\rarr B}$.
Note that achieving the quantum capacity, which may exceed the one-shot coherent information of a quantum channel $I_{\rm c}(\cN_1^{A_1 \rarr C_1})$ due to its superadditivity, generally requires encoding and decoding collectively over $N$-block uses of the channel $\cN_1$.
On the other hand, in the case that unlimited pre-shared entanglement is available, the entanglement-assisted quantum capacity $Q_{\rm E}(\cN_1)$ is given by the mutual information of a quantum channel~\cite{bennett2002entanglement, wilde2013QItheory, bennett2014quantum}:
\begin{align}
\label{eq:entassist capacity thm}
    Q_{\rm E}(\cN_1) = \f{1}{2}I(\cN_1^{A_1 \rarr C_1}),
\end{align}
where $I(\cT^{A\rarr B}) = \max_{\ket{\rho}^{AA'}}\big[H\big(\cT^{A\rarr B}(\rho^A)\big) + H(\rho^A) - H\big(\cT^{A\rarr B}(\ketbra{\rho}{\rho}^{AA'})\big)\big]$ is the mutual information of a quantum channel $\cT^{A\rarr B}$~\cite{wilde2013QItheory, khatri2024principlemodern}.
The maximization is taken over all purifications $\ket{\rho}^{AA'}$ of the state $\rho^A$ with a reference system $A'$.
Unlike the coherent information of a quantum channel $I_{\rm c}(\cT^{A\rarr B})$, the mutual information of a quantum channel $I(\cT^{A \rarr B})$ is additive, and thus Eq.~\eqref{eq:entassist capacity thm} does not involve regularization via a limit over $N$.

From the discussion of the decoupling and the random encoding (see, e.g.,~\cite{klesse2007approxqecrandomcode, hayden2008decoupling, dupuis2010decoupling, dupuis2014one, watrous2018TheoryQI, khatri2024principlemodern}), it is known that, in the entanglement-non-assisted setting, when the achievable rate $\mathrm{R}$ is below the quantum capacity $Q(\cN_1)$, there exists a suitable isometric encoder that achieves decoupling with asymptotically vanishing error: $\epsilon \to 0$ as $N \to \infty$.
This is also the case in the entanglement-assisted setting with unlimited entanglement, where $\mathrm{R}$ must satisfy $\mathrm{R} < Q_{\rm E}(\cN_1)$.
Then, Proposition~\ref{prop: uhlmann's dec} implicitly provides a decoder under which the recovery error asymptotically tends to zero.
This guarantees the existence of a decoder that can be used to achieve the communication rate asymptotically approaching the quantum capacity; however, this does not provide an explicit procedure for constructing the decoder, and all details, including its computational cost, remain unclear\footnote{In recent work, explicit algorithms directly addressing the Uhlmann's theorem have also been explored; see Refs.~\cite{bostanci2023unicompuhlmann, utsumi2025algorithmsUhlmanntrans}.}.

Constructing a high-performance decoder explicitly in the form of quantum circuits poses significant challenges.
The decoupling approach offers guidance toward this goal, as the relation between the decoupling and the quantum capacity implies that by designing decoders under the decoupling condition, we obtain explicit decoders that can be used to asymptotically achieve a rate approaching the quantum capacity.
Hereafter, to construct explicit decoders, we investigate a general channel $\cN$ and the one-shot scenario. For discussing the quantum capacity, it suffices to apply the decoder to the case $\cN = \cN_1^{\otimes N}$ and consider the asymptotic limit.


\subsection{Petz recovery map}
\label{sec: petz prelim.}
One of the explicit decoders we may use is the Petz recovery map~\cite{petz1986sufficient, petz1988sufficiency}, which has been intensely studied~\cite{ng2010simpleapproach, wilde2013QItheory, beigi2016decoding, lauten2022approx}. The Petz recovery map is developed from a quantum analog of Bayes theorem based on the idea that there can be a reverse channel that counteracts the effect of noise. The general form of the Petz recovery map is determined by a map $\cT$ and a reference state $\sigma$, given by
\begin{align}
    &\mathcal{P}^{B \rarr A}_{\sigma,\cT}(\cdot)  
    = (\sigma^A)^{\f{1}{2}}(\cT^{A\rarr B})^\dag \Big([\cT(\sigma^A)]^{-\f{1}{2}} (\cdot) \notag\\
    &\hspace{7pc}[\cT(\sigma^A)]^{-\f{1}{2}}\Big)(\sigma^A)^{\f{1}{2}},
\end{align}
where $(\cT^{A\rarr B})^\dag$ is the adjoint map of $\cT^{A\rarr B}$ with respect to the Hilbert--Schmidt inner product. The Petz recovery map is composed of three CP maps:
\begin{align}
    &(\cdot) \rarr [\cT(\sigma^A)]^{-\f{1}{2}}(\cdot)[\cT(\sigma^A)]^{-\f{1}{2}}, \label{Eq:Petz1}\\
    &(\cdot) \rarr (\cT^{A \rarr B})^\dag(\cdot), \label{Eq:Petz2}\\
    &(\cdot) \rarr (\sigma^A)^{\f{1}{2}}(\cdot)(\sigma^A)^{\f{1}{2}}, \label{Eq:Petz3}
\end{align}
and achieves the perfect recovery for the reference state $\sigma^A$, i.e., $\cP^{B \rarr A}_{\sigma, \cT}(\cT^{A \rarr B}(\sigma^A))=\sigma^A$.

For the recovery error of the Petz recovery map, the following is known. 

\begin{proposition}[Barnum-Knill's theorem \cite{barnum2002reversing}]
    \label{prop: barnum knill}
For any state $\rho^A$ and any channel $\cT^{A \rarr B}$, it holds that 
    \begin{align}
        &\mathrm{F}\big(\rho^{AR}, \cP_{\rho, \cT}^{B \rarr A}\circ\cT^{A \rarr B}(\rho^{AR})\big) \notag \\
        &\hspace{0pc}\geq \big[\max_{\cR} \mathrm{F}\big(\rho^{AR}, \cR^{B \rarr A}\circ \cT^{A \rarr B}(\rho^{AR})\big)\big]^2,
    \end{align}
where $\rho^{AR} = \ketbra{\rho}{\rho}^{AR}$ is a purified state of $\rho^A$. The maximum is taken over all quantum channels $\cR^{B \rarr A}$.
\end{proposition}
This proposition guarantees that if there exists a decoder that recovers information with a small error, the Petz recovery map also recovers it with a small error.

\begin{figure}
    \centering
    \includegraphics[width=80mm]{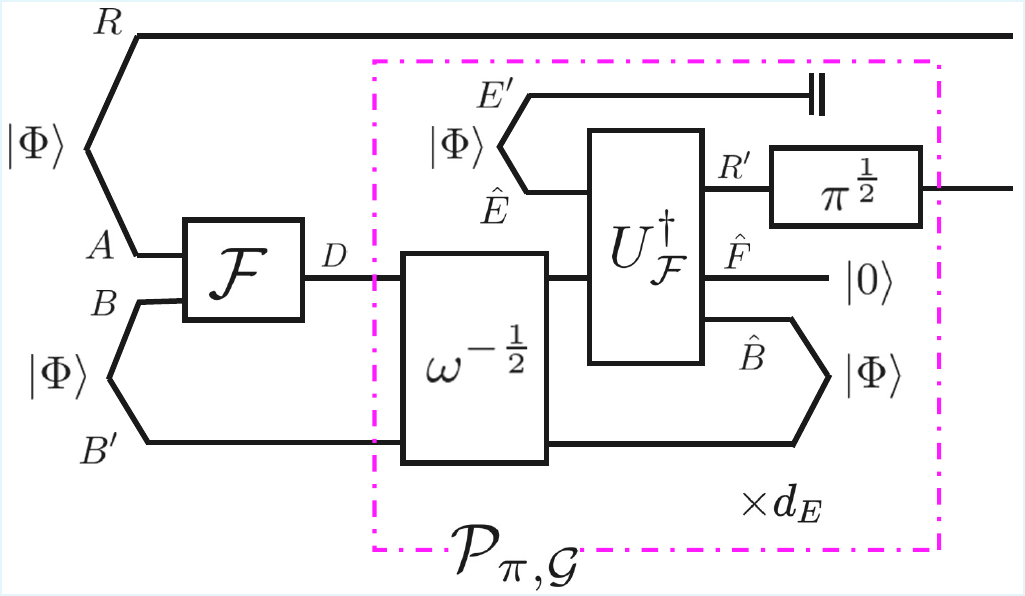}
    \caption{\footnotesize A diagram of the Petz recovery map applied to our setting. 
    The dash-dotted box corresponds to the Petz recovery map $\cP_{\pi, \cG}$ given in Eq.~\eqref{Eq:PetzOurSetting}.
    The boxes of $(\omega^{DB'})^{-1/2}$ and $(\pi^{R'})^{1/2}$ represent that $(\cdot) \rarr (\omega^{DB'})^{-1/2}(\cdot)(\omega^{DB'})^{-1/2}$ and $(\cdot) \rarr (\pi^{R'})^{1/2}(\cdot)(\pi^{R'})^{1/2}$, respectively. The double vertical lines represent that the qubits of that system are traced out.}
    \label{fig: original Petz}
\end{figure}

To apply the Petz recovery map to our setting, let $F$ be the system such that $ABF = ED$, and a unitary $U_\cF^L$ be defined by 
\begin{equation}
\label{eq: stinespring iso-uni}
    V_\cF^{AB \rarr ED} = U_\cF^L\ket{0}^F, 
\end{equation}
where $L=ABF = ED$. Using this unitary, Eq.~\eqref{eq: stine dilation by iso} is rephrased as
\begin{equation}
\label{eq: stinespring unitary of F}
    \cF^{AB \rarr D}(\cdot) = \tr_E\big[U_\cF^L(\cdot\otimes\ketbra{0}{0}^F)(U_\cF^L)^\dag\big].
\end{equation}
We use $\cG^{A \rarr DB'}(\cdot) \coloneqq \cF^{AB \rarr D}(\cdot \otimes  \Phi^{BB'})$ and fix the reference state to be the CMS $\pi^A$. The explicit form of the Petz recovery map in our setting is then given by
\begin{align}
    &\cP_{\pi, \cG}^{DB'\rarr R'}(\omega^{RDB'})   \notag\\
    &= d_E\tr_{E'}\Big[(\pi^{R'})^{1/2}\bra{\Phi}^{\hat{B}B'}\bra{0}^{\hat{F}} (U_\cF^{\hat{L}})^\dag   \notag\\
    &\hspace{3pc}\big[(\omega^{DB'})^{-1/2}\omega^{RDB'}(\omega^{DB'})^{-1/2}\otimes \Phi^{\hat{E}E'}\big] \notag\\
    &\hspace{5pc}U_\cF^{\hat{L}}\ket{\Phi}^{\hat{B}B'}\ket{0}^{\hat{F}}(\pi^{R'})^{1/2}\Big], \label{Eq:PetzOurSetting}
\end{align}
where $\hat{L}$ is equal to $R'\hat{F}\hat{B} = \hat{E}D$. See also the diagram in Fig.~\ref{fig: original Petz}.

By combining Proposition~\ref{prop: barnum knill} with Proposition~\ref{prop: uhlmann's dec} and the Fuchs--van de Graaf inequalities Eq.~\eqref{eq: fuchs van de graaf}, we derive the following statement, which relates the recovery error of the Petz recovery map $\cP_{\pi, \cG}^{DB' \rarr R'}$ against $\cF^{AB \rarr D}$ to the decoupling condition: if there exists a state $\tau^E$ such that $\|\omega^{RE} - \pi^R \otimes \tau^E\|_1 \leq \epsilon$, then the recovery error of the Petz recovery map in the above setting is given by
    \begin{equation}
    \label{cor: recovery error of Petz}
        \Delta(\cP_{\pi, \cG}|\cF) \leq 2\epsilon^{1/4}.
    \end{equation}
As discussed in \ref{sec: decoupling and Uhlmann}, the decoupling is asymptotically achieved by an appropriately chosen encoder. Since the upper bound on the recovery error of $\cP_{\pi, \cG}$ also asymptotically tends to zero with such an encoder, its communication rate can approach the quantum capacity. 

Although it is not clear from the definition how the Petz recovery map can be implemented by quantum circuits, an algorithmic implementation was provided in~\cite{gilyen2022petzmap}. The algorithm is based on the fact that the Petz recovery map is a CPTP map as a whole, which allows its implementation through the direct use of the QSVT. However, its circuit complexity grows exponentially with the number of qubits due to the full implementation of the CP maps in Eqs.~\eqref{Eq:Petz1} to~\eqref{Eq:Petz3}.


\subsection{Two-step construction of a decoder for the Hayden--Preskill protocol}

In~\cite{yoshida2017efficient}, a decoding circuit was provided for recovering quantum information in the HP protocol~\cite{hayden2007black}. The HP protocol formulates the information paradox of black holes based on the qubit-erasure noise with a restriction that the encoding operation is given by a random unitary dynamics. That is, the encoder $\cE$ and the noise $\cN$ in Fig.~\ref{fig: ent asist} are given by a random unitary and the partial trace over a subsystem $E$ of $C$, respectively, where $AB = C$.

The construction of a decoder consists of two steps.
The first step is to construct a decoding protocol with post-selection. This is achieved by ``emulating'' the dynamics of the encoding unitary and the erasure noise in the receiver's local system and teleporting the output of the noise by performing the measurement in a maximally entangled basis. More specifically, after preparing the emulated systems in the receiver's local system, the receiver measures the output of the noise and the corresponding emulated output in the maximally entangled basis. If a desired outcome is obtained, the emulated output becomes as if it were in the same quantum state as that of the noisy output. In this case, the effect of the erasure noise is canceled by the dynamics emulated in advance in the local system, and the receiver succeeds in recovering quantum information. Note that this protocol does not succeed with certainty as it requires post-selection.

In the second step of the construction, this post-selection is removed by replacing the measurement with the AA algorithm. This replacement can be understood as aiming to amplify the probability of obtaining the desired outcome.
As a result, a decoding quantum circuit for the HP protocol without post-selection is constructed.

The reason for this construction to work, or more specifically the AA algorithm to work, strongly relies on the specific properties of the HP protocol, as we will later elaborate on.
The proof technique is also fully tailored to the HP protocol and cannot be simply generalized to other noise models.
Hence, this decoding strategy with a standard AA algorithm for the HP protocol cannot be directly applied to general noisy situations.


\subsection{Various amplitude amplification protocols}
\label{sec: explanation QSVT-based FPAA}

The AA algorithm is a common technique to enhance the measurement probability to obtain a desired output, and it provides a quadratic speedup over classical algorithms. Various amplitude amplification protocols have been proposed so far, and each has different features.

To remind the standard AA algorithm, let an initial state and a desired state be $\ket{\psi}$ and $\ket{\xi}$, respectively. We consider iteratively applying unitaries $I - 2\ketbra{\xi}{\xi}$ and $I - 2\ketbra{\psi}{\psi}$, $t$ times to the initial state. This iterative application approximately achieves the state transformation, such as
\begin{equation}
\label{Eq:standardAA}
    \ket{\psi} \mapsto \ket{\xi}, \ \ \text{if $t = \big\lfloor\pi/(4|\braket{\xi}{\psi}|)$}\big\rfloor.
\end{equation}
One feature of this standard AA algorithm is that this desired state is not a fixed point of the operation. That is, if the number $t$ of iterations exceeds the value in Eq.~\eqref{Eq:standardAA}, the resulting state becomes different from $\ket{\xi}$. This is known as an \emph{overcook} problem~\cite{brassard1997searching}. For this reason, it is crucial to precisely know the value of $|\braket{\xi}{\psi}|$ in advance.

The overcook problem is circumvented by the FPAA~\cite{yoder2014fixed}. The FPAA algorithm also consists of the iterations of unitaries, but there is a threshold number $t_{\rm th}$ of iterations such that 
\begin{equation}
\label{Eq:FPAA}
    \ket{\psi} \mapsto e^{i\chi}\ket{\xi},\ \  \forall t \geq t_{\rm th} = \cO(1/|\braket{\xi}{\psi}|),
\end{equation}
is approximately achieved, where $\chi$ is an unknown phase. Unlike the standard AA algorithm, if $t \geq t_{\rm th}$, the state always stays in this form. As the unknown phase $\chi$ is typically not important in many applications of the AA algorithm, the FPAA should resolve the overcook problem. Importantly, the FPAA works if a lower bound on $|\braket{\psi}{\xi}|$ is known in advance, and precise estimation of the value is not necessary.

While the FPAA may be sufficient in most applications, one can get rid of the unknown phase $\chi$ by the QSVT-based FPAA~\cite{gilyen2019qsvt, Gilyn2019thesis, martyn2021grand}. We will describe the QSVT-based FPAA in more detail in the next section. The QSVT-based FPAA iterates certain unitaries $t$ times, and one can approximately achieve
\begin{equation}
\label{Eq:QSVTbasedAA}
    \ket{\psi} \mapsto \ket{\xi},\ \  \forall t \geq t_{\rm th} = \cO(1/|\braket{\xi}{\psi}|),
\end{equation}
without any unknown phases that could be problematic depending on the goals.
As pointed out in~\cite{Gilyn2019thesis}, this is one of the unique features of the QSVT-based FPAA, which may help in achieving tasks that are not achievable by any other AA-type algorithms.
Constructing explicit examples of such tasks, especially those of practical importance or those contributing to existing problems, is of theoretical interest.


\subsection{Quantum singular value transformation-based fixed-point amplitude amplification}
\label{sec: explain QSVT FPAA}

Our construction of decoders employs the QSVT-based FPAA. We here introduce it briefly. For further details, see, e.g.,~\cite{gilyen2019qsvt, Gilyn2019thesis, martyn2021grand}.

We first introduce a notion of block-encoding~\cite{Low2019HamiltonianQubitize, gilyen2019qsvt}.
When a certain operator of interest $\Lambda$ such that $\|\Lambda\|_\infty \leq 1$ is encoded as a block of a larger unitary $U$:
\begin{align}
\label{inteq:5}
        U
     = \hspace{1mm}
\begin{blockarray}{ccc}
&\hspace{1.5mm} \Pi_1 &  & \vspace{1mm}\\
\begin{block}{c(cc)}
   \Pi_2  & \Lambda & \hspace{0mm} \cdot \hspace{2mm}\\
     & \cdot & \hspace{0mm} \cdot \hspace{2mm}\\
\end{block}
\end{blockarray}
\   \longleftrightarrow  \  \Lambda = \Pi_2 U \Pi_1,
\end{align}
the unitary $U$ is called a block-encoding of $\Lambda$.
Here, $\Pi_1$ and $\Pi_2$ are projectors which specify the top-left block of $U$.
The normalization $\|\Lambda\|_\infty \leq 1$ ensures that $U$ is a unitary with appropriate choices of the remaining blocks of $U$.

The QSVT~\cite{gilyen2019qsvt, Gilyn2019thesis} is a powerful technique to transform $\Lambda$ to its even or odd polynomial function, by multiple uses of $U$ and $U^\dag$. 
In this study, we only focus on degree-$t$ real odd polynomials $Q_{t, \phi}$ that satisfy $|Q_{t, \phi}(x)| \leq 1$ for all $x \in [-1, 1]$, where $\phi = (\phi_1, \phi_2, \ldots, \phi_t)$ is a sequence of parameters with $\phi_j \in (-\pi, \pi]$ for $j = 1, 2, \ldots, t$, which determines the form of the polynomial $Q_{t,\phi}$.

The QSVT with a real polynomial is such that it uses $U$ and $U^\dag$ a total of $2t$ times to perform the transformation:
\begin{align}
        &U
     = \!\hspace{1mm}
\begin{blockarray}{ccc}
&\hspace{1.5mm} \Pi_1 &  & \vspace{1mm}\\
\begin{block}{c(cc)}
  \Pi_2 & \Lambda & \hspace{0mm} \cdot \hspace{2mm}\\
    & \cdot & \hspace{0mm} \cdot \hspace{2mm}\\
\end{block}
\end{blockarray} \\
\label{inteq:6}
&\overset{\rm QSVT}{\longrightarrow}  \
        G_{t, \phi}
     = \!\hspace{1mm}
\begin{blockarray}{ccc}
& \Pi_1' &  & \vspace{1mm}\\
\begin{block}{c(cc)}
  \Pi_2' & Q_{t, \phi}(\Lambda) \otimes \ketbra{0}{0} & \hspace{0mm} \cdot \hspace{2mm}\\
    & \cdot & \hspace{0mm} \cdot \hspace{2mm}\\
\end{block}
\end{blockarray}\hspace{1mm},
\end{align}
where $\Pi_m' = \Pi_m \otimes \ketbra{0}{0}$ for $m = 1, 2$, the projectors with an ancilla qubit.
When $Q_{t, \phi}$ is odd, its action on a matrix is defined by
\begin{align}
\label{eq:P poly def}
    Q_{t, \phi}(\Lambda) =
    \sum_\mu Q_{t, \phi}(s_\mu)\ketbra{\xi_\mu}{\psi_\mu},
\end{align}
where $\Lambda = \sum_\mu s_\mu \ketbra{\xi_\mu}{\psi_\mu}$ is the singular value decomposition of $\Lambda$.
The unitary $G_{t, \phi}$ is constructed from $U$ as follows. 
Let $W_m(\theta) \coloneqq e^{i\theta(2\Pi_m -\mathbb{I})}$ for $m = 1, 2$ and $\theta \in (-\pi, \pi]$, and define $W_{t, \phi}$ as 
\begin{equation}
    W_{t, \phi} \coloneqq W_2(\phi_t)U\prod_{j = 1}^{(t-1)/2}W_1(\phi_{2j})U^\dag W_2(\phi_{2j-1}) U.
\end{equation}
Then, $G_{t, \phi}$ is given by
\begin{equation}\begin{split}
    &G_{t, \phi}\coloneqq W_{t, \phi}\otimes \ketbra{+}{+} + W_{t, -\phi} \otimes \ketbra{-}{-}.
\end{split}\end{equation}
The unitary $G_{t, \phi}$ constructed in this way satisfies
\begin{align}
\label{inteq:7}
    (\Pi_2 \otimes \bra{0}) G_{t, \phi} (\Pi_1\otimes\ket{0}) = Q_{t, \phi}(\Lambda),
\end{align}
as indicated by Eq.~\eqref{inteq:6}.

Note that, to implement the QSVT in practice, one needs to compute the sequence of phases $\phi$ corresponding to a given $Q_{t, \phi}$. It is known that we can compute it in $\cO(\mathrm{poly}(t))$ time even by a classical computer~\cite{Haah2019product, Chao2020FindingAF, dong2021findingphase, Lin2022LectureNotes, Mizuta2023vzw}.

By choosing a polynomial $Q_{t, \phi}$ appropriately, we can implement various useful quantum algorithms within the QSVT framework.
In this work, we particularly consider a real odd polynomial that closely approximates the sign function:
\begin{equation}
\label{eq: sign func.}
    \mathrm{sign}(x) = \left\{
    \begin{array}{ll}
         1&  ( x > 0 ) \\
         0&  ( x = 0 ) \\
         -1& ( x < 0 )\hspace*{1mm}.
    \end{array}
    \right.
\end{equation}
This case is referred to as the QSVT-based FPAA, also known as the singular vector transformation~\cite{gilyen2019qsvt, Gilyn2019thesis}.
The following lemma ensures that there exists a polynomial $Q_t^{\rm sign}$ that well approximates the sign function.

\begin{lemma}[Polynomial approximation of the sign function \cite{Low2017HamiltonianSB, gilyen2019qsvt, Gilyn2019thesis, Lin2020nearoptimalground, martyn2021grand, martyn2023efficient, Mitarai2023perturbationtheory, toyoizumi2024hamiltonian}]
\label{lem: poly apprx sign func}
    For $\delta \in (0, 1/2)$ and $\beta \in (0, 1)$, and for any odd integer $t \geq \big\lceil\frac{8e}{\beta}\log(2/\delta)\big\rceil$, there exists a real polynomial $Q_t^{\rm sign}(x)$ of degree $t$ such that
    \begin{itemize}
        \item $|Q_t^{\rm sign}(x)| \leq 1$ for $x \in [-1, 1]$,
        \item $|Q_t^{\rm sign}(x) - \mathrm{sign}(x)| \leq \delta$ for $x \in [-1, -\beta]\cup[\beta, 1]$.
    \end{itemize}
\end{lemma} 

To perform the QSVT-based FPAA, we take the phase sequence $\phi = (\phi_1, \ldots, \phi_t)$ in the QSVT so that the polynomial $Q_{t, \phi}$ becomes $Q_t^{\rm sign}$. With this polynomial, the matrix $\Lambda$ is mapped to $Q_t^{\rm sign}(\Lambda) \approx \sign(\Lambda) = \sum_\mu \ketbra{\xi_\mu}{\psi_\mu}$.
We thus obtain an operation that transforms the right singular vectors $\ket{\psi_\mu}$ to the corresponding left singular vectors $\ket{\xi_\mu}$, without introducing any unknown phases.

The range of the degree $t$ of $Q_t^{\rm sign}$ is determined as follows. 
Let $\gamma_{\delta, \beta}$ be given by
\begin{align}
    &\gamma_{\delta, \beta} = 2 \Big\lceil\max\Big\{
    \frac{e}{2\beta} \sqrt{W\Big(\frac{8}{\pi\delta^2}\Big) W\Big(\frac{512}{e^2\pi\delta^2}\Big)}, \notag\\
    &\hspace{4pc}\sqrt{2}\,W\Big(\frac{4}{\beta\delta}\sqrt{\frac{2}{\pi}W\Big(\frac{8}{\pi\delta^2}\Big)}\Big)\Big\}\Big\rceil + 1,
\end{align}
with the Lambert $W$ function $W(x)$~\cite{Corless1996onthelambert}.
Then, for any odd integer $t \geq \gamma_{\delta, \beta}$, there is a degree-$t$ polynomial that approximates $\sign(x)$ within error $\delta$ over $|x| \in [\beta, 1]$~\cite{martyn2023efficient}.
Since $W(x) \leq \log x$ holds for $x > e$~\cite{Hassani2005Approximationlambert, Hoorfar2008inequlambert}, we obtain $\gamma_{\delta, \beta} \leq \big\lceil\frac{8e}{\beta}\log(2/\delta)\big\rceil$ for $\delta \in (0, 1/2)$ and $\beta \in (0,1)$. 
As a result, it suffices to set $t$ to an odd integer satisfying $t \geq \big\lceil\frac{8e}{\beta}\log(2/\delta)\big\rceil$.

\section{Main results}
\label{sec: main results}

In this section, we summarize our results. We provide explicit quantum circuit constructions of two decoders and evaluate their performance. 
The generalized YK decoder is presented in \ref{sec: GYK decoder}, and the Petz-like decoder in \ref{sec: IPM decoder}. 
We investigate the complexity of the decoders in \ref{sec: comparison} and \ref{sec: specific compare}.

Both decoders are constructed by the extended two-step construction. We first provide a protocol with post-selection and then transform the protocol into the one without post-selection.
To achieve the latter, we use the QSVT-based FPAA algorithm, which is crucial for circumventing issues that arise if other AA-type algorithms are used.

\subsection{Generalized YK decoder}
\label{sec: GYK decoder}

In \ref{sec: prob GYK}, we investigate a decoding protocol with post-selection that works for general encoding maps and noisy channels.
We then show in \ref{sec: det GYK} that the protocol can be lifted up to a decoder, by using the QSVT-based FPAA.


\subsubsection{Decoding protocol with post-selection}
\label{sec: prob GYK}

\begin{figure} 
    \centering
    \includegraphics[width=80mm]{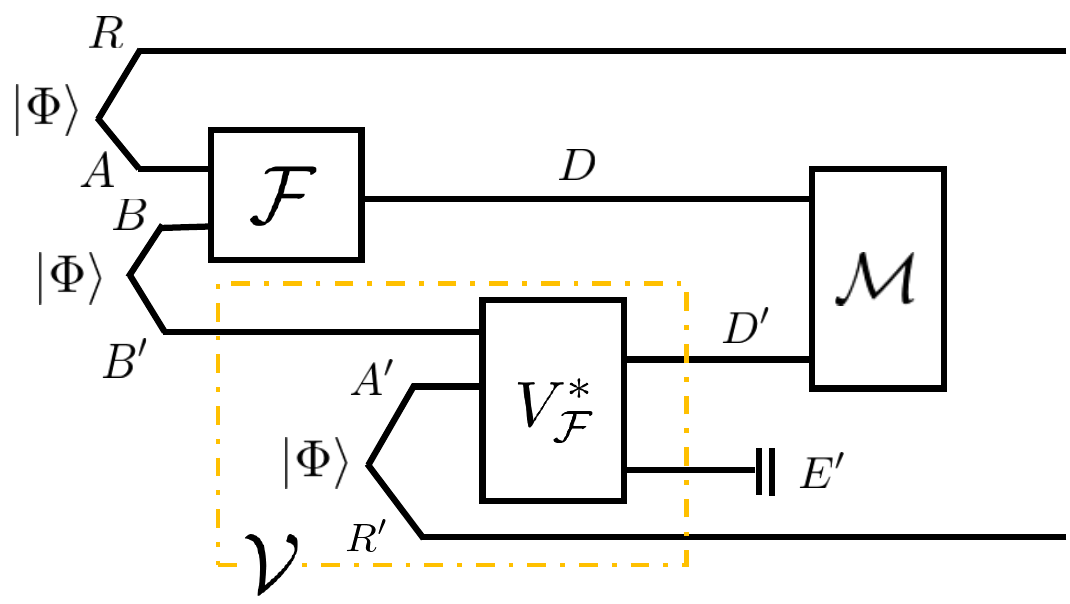}
    \caption{A diagram of the protocol with post-selection for the generalized YK decoder. The double vertical lines represent that the qubits of that system are traced out. The dash-dotted box corresponds to the isometry map $\cV^{B' \rightarrow D'E'R'}$ defined in Eq.~\eqref{eq: R cptp map}.}
    \label{fig: prob GYK}
\end{figure}

The decoding protocol with post-selection consists of the following three steps. 
See Fig.~\ref{fig: prob GYK} as well.
\begin{enumerate}
    \item The receiver prepares ancilla qubits in the system $A'R'$, and then generates a MES $\Phi^{A'R'}$, which is regarded as a copy of the MES $\Phi^{AR}$.
    \item The receiver applies an isometry $(V_\cF^{A'B'\rarr E'D'})^*$ onto $A'B'$, where $V_\cF^{AB \rarr ED}$ is a Stinespring isometry of $\cF^{AB \rarr D}$ and $E$ is an environment of the channel $\cF$. The complex conjugate is taken in the computational basis.
    \item The receiver performs a binary measurement $\mathcal{M} \coloneqq \{\ketbra{\Phi}{\Phi}^{DD'},  \mathbb{I}^{DD'}  - \ketbra{\Phi}{\Phi}^{DD'}\}$ on $DD'$. When the former result of the measurement $\cM$ is obtained, this protocol succeeds.
\end{enumerate}
Note that, in this protocol, all the systems with a prime, i.e., $A'$, $B'$, $R'$, $D'$, and $E'$, in addition to the output system $D$ of the channel $\cF$ are in the hands of the receiver. Thus, the above protocol can be executed by the receiver.

The Stinespring dilation $V_\cF^{AB \rarr ED}$ in the step~2 is not uniquely determined from a given channel $\cF^{AB \rarr D}$; the dilation has the freedom of applying additional isometries on the environment $E$. However, the protocol works for any choice of $V_\cF^{AB \rarr ED}$. The receiver can choose an arbitrary Stinespring dilation of the channel $\cF^{AB \rarr D}$.

For future use, we denote the operation up to the step~2 of the above protocol by an isometry map $\cV^{B'\rarr D'E'R'}$. That is,
\begin{equation}\begin{split}
\label{eq: R cptp map}
    &\cV^{B' \rarr D' E'  R'}(\cdot) \\
    &\coloneqq  (V_\cF^{A'B' \rarr E'D'})^* (\cdot \otimes   \Phi^{A'R'})(V_\cF^{A'B' \rarr E'D'})^{\mathsf{T}}. 
\end{split}\end{equation}
We denote by $p_{\rm succ}$ and $\zeta_{\rm succ}$ the success probability and the output state after the success of $\cM$ in the step~3, respectively. The reduced state on $RR'$ of $\zeta_{\rm succ}$ is given~by 
\begin{equation}\begin{split}
&\zeta_{\rm succ}^{RR'} \\
&= \tr_{DD'E'}\Big[ \frac{1}{p_{\rm succ}}\ketbra{\Phi}{\Phi}^{DD'}  \cV^{B' \rarr D'E'  R'}(\omega^{RDB'})\Big].
\end{split}\end{equation}

In \ref{sec: proof of prob GYK}, we compute $p_{\rm succ}$ and the fidelity between $\zeta_{\rm succ}^{RR'}$ and $\Phi^{RR'}$, and then obtain
\begin{align}
    &p_{\rm succ} = \f{d_B}{d_D}2^{-H_2(RE)_{\omega}}, 
    \label{eq: probability of gyk}\\
    &\mathrm{F}\big(\zeta_{\rm succ}^{RR'}, \Phi^{RR'}\big) = \frac{1}{d_A} 2^{H_2(RE)_{\omega} - H_2(E)_{\omega}}.
    \label{eq: post select GYK}
\end{align}
The second equation, Eq.~\eqref{eq: post select GYK}, implies that if $\omega^{RE}$ decouples as $\omega^{RE} \approx \pi^R \otimes \omega^E$, the fidelity after post-selection becomes $\mathrm{F}\big(\zeta_{\rm succ}^{RR'}, \Phi^{RR'}\big) \approx 1$. In other words, the MES can be successfully recovered if the measurement is successful and the decoupling condition is sufficiently satisfied.
However, the success probability $p_{\rm succ}$ is exponentially small, even when $\omega^{RE}$ decouples. This means that the decoding protocol with post-selection fails in most cases.


\subsubsection{Construction of the generalized YK decoder}
\label{sec: det GYK}

We now consider upgrading the decoding protocol with post-selection to a decoder without post-selection by the QSVT-based FPAA algorithm.
Let us first outline how this could be achieved.

Compared to the standard situations of using AA-type algorithms, we need a more careful analysis since we can manipulate only a part of the system. 
To clarify this point, we denote by $\ket{\omega_0}^{REDD'E'R'}$ a purified state after the step 2, that is, we purify the state before the measurement in Fig.~\ref{fig: prob GYK} by an environment $E$ of the channel $\cF$. We may divide this state into $RE$, which we have no access to, and $DD'E'R'$, which we can manipulate. This leads to the Schmidt decomposition: 
\begin{equation}
    \ket{\omega_0}^{REDD'E'R'} = \sum_\mu \sqrt{\lambda_\mu}\ket{\eta_\mu}^{RE}\ket{\psi_\mu}^{DD'E'R'}, \label{Eq:omega_0_first_to_appear}
\end{equation}
with Schmidt coefficients $\{\sqrt{\lambda_\mu}\}_\mu$, and orthonormal bases $\{ \ket{\eta_\mu}^{RE}\}_\mu$ and $\{\ket{\psi_\mu}^{DD'E'R'}\}_\mu$ in $RE$ and $DD'E'R'$, respectively.
By applying an AA-type algorithm to $DD'E'R'$ of $\ket{\omega_0}^{REDD'E'R'}$, we aim to achieve the transformation:
\begin{equation}
    \label{eq: trans of the schmidt basis in FPAA}
    \ket{\psi_\mu}^{DD'E'R'} \mapsto \ket{\xi_\mu}^{DD'E'R'},
\end{equation}
for each $\mu$, where $\{\ket{\xi_\mu}^{DD'E'R'}\}_\mu$ is the Schmidt basis of the post-selected state after the measurement in the step~3. As it is post-selected by the MES $\ketbra{\Phi}{\Phi}^{DD'}$ and due to the symmetry of the state, we can show that 
\begin{equation}
    \ket{\xi_\mu}^{DD'E'R'} = \ket{\Phi}^{DD'}\ket{\eta_\mu^*}^{E'R'},
\end{equation}
where the complex conjugate acts on the coefficients when the state is expanded in the computational basis.
Hence, if we can achieve the transformation given by Eq.~\eqref{eq: trans of the schmidt basis in FPAA} for all $\mu$ simultaneously while maintaining the superposition, the entire state is transformed as 
\begin{align}
    &\ket{\omega_0}^{REDD'E'R'} \mapsto \ket{\Phi}^{DD'} \sum_\mu \sqrt{\lambda_\mu}\ket{\eta_\mu}^{RE}\ket{\eta_\mu^*}^{E'R'}. \label{Eq:omega_0_to_something}
\end{align}
Assuming the decoupling between $R$ and $E$, we can further show that this is close to $\ket{\Phi}^{DD'} \ket{\tau}^{EE'} \ket{\Phi}^{RR'}$.
This statement is non-trivial, and we use the Powers--St\o rmer inequality~\cite{powers1970free, kittaneh1987inequalities} to prove it.
As a result, we obtain the MES $\ket{\Phi}^{RR'}$ between the reference $R$ and the subsystem $R'$ in the receiver's hands, completing the recovery of quantum information. See~\ref{sec: proof of det GYK} for the details.

The remaining and crucial question is how we could simultaneously achieve the transformation in Eq.~\eqref{eq: trans of the schmidt basis in FPAA} for all $\mu$. The standard AA fails to achieve this because the number of iterations of operations is sensitive to the exact value of $|\braket{\xi_\mu}{\psi_\mu}|$ (see Eq.~\eqref{Eq:standardAA}), which differs for each $\mu$ in general. Thus, although we could achieve Eq.~\eqref{eq: trans of the schmidt basis in FPAA} for some $\mu$, other states will be overcooked or undercooked, which ends up in a failure of achieving the transformation in Eq.~\eqref{Eq:omega_0_to_something}.
Note that this issue does not arise if all the inner products $|\braket{\xi_\mu}{\psi_\mu}|$ are almost the same. This is the case for the HP protocol and is why the original decoding protocol~\cite{yoshida2017efficient} for the HP protocol works with the standard AA algorithm.

An issue still arises even if we use the original FPAA~\cite{yoder2014fixed}. Although the overcook and undercook problems can be circumvented (see Eq.~\eqref{Eq:FPAA}), it eventually results in
\begin{equation}
    \ket{\Phi}^{DD'}\sum_\mu e^{i\chi_\mu} \sqrt{\lambda_\mu}\ket{\eta_\mu}^{RE}\ket{\eta_\mu^*}^{E'R'}.
\end{equation}
Again, this fails to achieve Eq.~\eqref{Eq:omega_0_to_something}.
This FPAA does not work since we need to operate on the state in a superposition, making the unknown phase $\chi_\mu$ a relative one. 
The phase $\chi_\mu$ generally depends on the overlap between the input and output states, $\braket{\xi_\mu}{\psi_\mu}$, and computing it would not be straightforward.
This concern was pointed out in~\cite{Gilyn2019thesis} as a general remark.

For these reasons, the only AA-type algorithm that we can employ to achieve Eq.~\eqref{Eq:omega_0_to_something} is the one based on the QSVT, which in fact works well for our purpose (see Eq.~\eqref{Eq:QSVTbasedAA}).
We emphasize here that the reason why only the QSVT-based FPAA works is that, in the decoding task, we have access only to one part of the entangled systems. In this situation, the superposition makes the relative phases, that arise from other AA-type algorithms,
harmful and results in the failure of decoding.
Hence, our decoding protocol can be considered as a concrete and practical example that highlights the unique feature of the QSVT-based FPAA compared to other AA-type algorithms.

We now explain a concrete algorithm for constructing a decoder.
Let $G_{t, \phi}$ be the unitary corresponding to the QSVT-based FPAA, where $t\in \mathbb{N}$ and $\phi=(\phi_1, \phi_2, \ldots, \phi_t) \in (-\pi, \pi]^t$ are the parameters of the algorithm. We replace the step 3 in the previous section with the application of the unitary $G_{t, \phi}$. 
\begin{enumerate}
    \item[3'.] The receiver prepares an auxiliary single-qubit state $\ket{0}^H$ in a system $H$, and then applies a unitary $G_{t, \phi}^{DD'E'R'H}$,
    with appropriate $t$ and $\phi$ to approximate the sign function.
\end{enumerate}
The details of the unitary $G_{t, \phi}$ will be explained later in this section.

\begin{figure*}
    \centering
    \includegraphics[width=160mm]{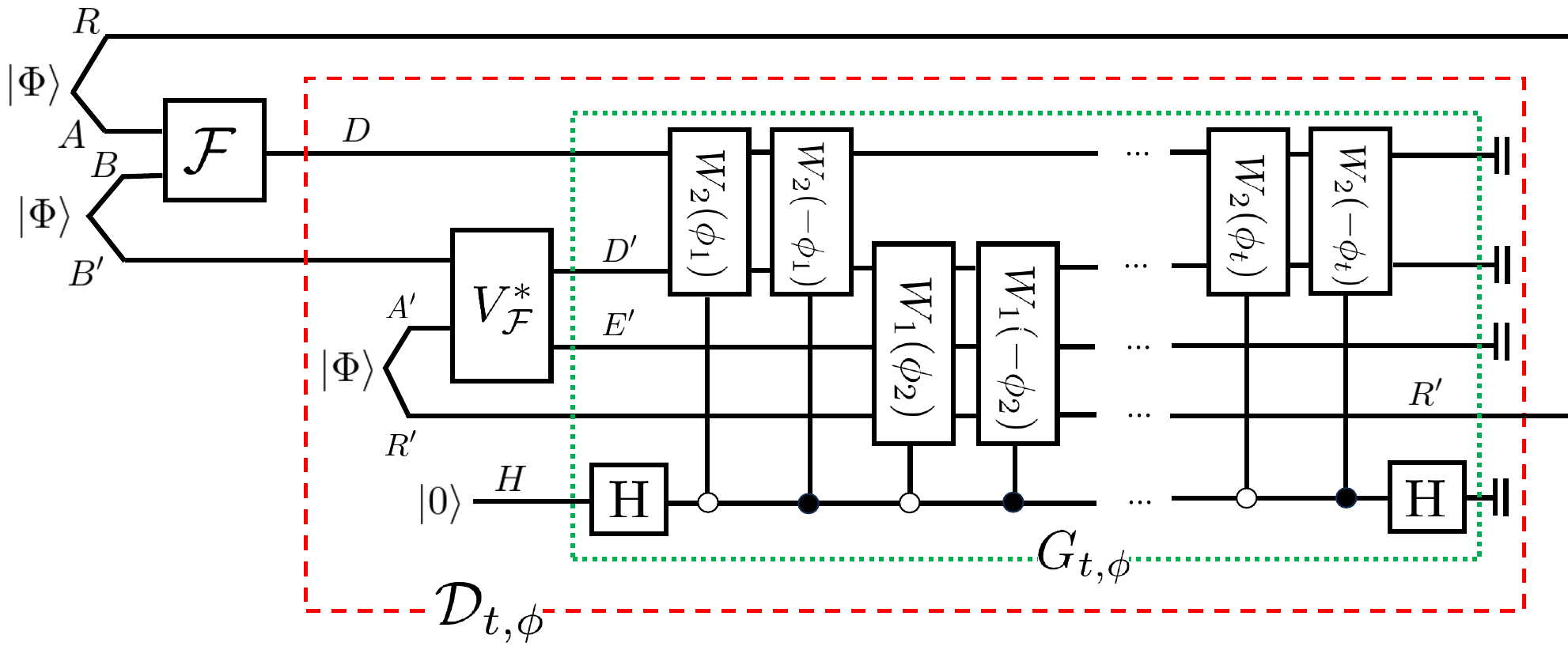}
    \caption{A diagram of the generalized YK decoder. Open circles imply that the gates are controlled by $\ket{0}$, while closed circles indicate the ones controlled by $\ket{1}$. The gate $\mathrm{H}$ is the single-qubit Hadamard gate. The red dashed and green dotted boxes correspond to the generalized YK decoder $\cD_{t, \phi}$ defined in Eq.~\eqref{eq: def. of gyk decoder}, and the unitary $G_{t, \phi}$ by the QSVT-based FPAA algorithm given in Eq.~\eqref{eq: FPAA unitary}, respectively. The unitary $W_m(\theta)$ ($m=1, 2$) is defined in Eq.~\eqref{eq: rotation unitary W}.}
    \label{fig: det GYK}
\end{figure*}

This replacement allows us to obtain a decoder without post-selection, i.e., the generalized YK decoder. All together, the decoding CPTP map is given by
\begin{align}
\label{eq: def. of gyk decoder}
&\mathcal{D}_{t,\phi}^{DB' \rarr R'}(\cdot) \notag\\
&\hspace{0.5pc}\coloneqq \tr_{DD'E'H}\big[G_{t, \phi}^{DD'E'R'H} \big(\cV^{B' \rarr D'E'R'}(\cdot) \notag\\
&\hspace{6pc}\otimes \ketbra{0}{0}^H\big)(G_{t, \phi}^{DD'E'R'H})^\dag\big], 
\end{align}
where $\cV^{B' \rarr D'E'R'}$ is defined in Eq.~\eqref{eq: R cptp map}. See Fig.~\ref{fig: det GYK} as well.

This construction of the generalized YK decoder is an extension of the original one for the HP protocol. However, it is not straightforward to analyze its decoding performance, as the original analysis is based on the specific details of the HP protocol and cannot be applied to any other noise. By developing a proof technique that relies on the symmetric structure of the generalized YK decoder, we prove the following theorem.

\begin{theorem}[Performance of the generalized YK decoder]
\label{thm: the gYK}
For a channel $\cF^{AB \rarr D}$, let $\bar{\cF}^{AB \rarr E}$ be its complementary channel, $\omega^{RE}$ be the state given by
\begin{equation}
    \omega^{RE} = \bar{\cF}^{AB \rarr E}(\Phi^{AR} \otimes \pi^B),
\end{equation}
and $\lambda_{\rm min}(\omega^{RE})$ be the non-zero minimum eigenvalue of $\omega^{RE}$.
Suppose that there exists a state $\tau^E$ such that $\|\omega^{RE} - \pi^R \otimes \tau^E\|_1 \leq \epsilon$. Then, for any $\delta \in (0, 1)$ and any odd integer $t$ satisfying
\begin{equation}
\label{eq: condition for t GYK}
    t \geq \bigg\lceil16e\sqrt{\f{d_D}{d_B\lambda_{\rm min}(\omega^{RE})}}\log(2/\delta)\bigg\rceil,
\end{equation}
there exist $\phi = (\phi_1, \phi_2, \ldots, \phi_t) \in (-\pi, \pi]^{ t}$ such that 
the recovery error $\Delta(\mathcal{D}_{t, \phi}|\cF)$ of the generalized YK decoder $\mathcal{D}_{t, \phi}^{DB' \rarr R'}$ is given by
    \begin{equation}
        \Delta(\mathcal{D}_{t, \phi}|\cF) \leq \sqrt{\epsilon} + \delta, \label{Eq:ErrorGYK}
    \end{equation}
and the circuit complexity of $\mathcal{D}_{t, \phi}^{DB' \rarr R'}$ is 
\begin{equation}
\label{eq: comp gYK}
    \cC\big(\mathcal{D}_{t, \phi}\big)   = \cO\Big(t  \big(\cC(U_\cF) + \log(d_D^2d_E/d_B)\big)\Big),
\end{equation}
and $\cO\big(\log(d_D^2d_E/d_B)\big)$ ancilla qubits suffice.
Here, $\cC(U_\cF)$ is circuit complexity of a unitary $U_\cF^L$ such that $U_\cF^L\ket{0}^F$ is a Stinespring isometry of $\cF^{AB \rarr D}$, and $L = ABF = ED$.
\end{theorem}

Theorem~\ref{thm: the gYK} shows in Eq.~\eqref{Eq:ErrorGYK} that the recovery error is dependent on $\epsilon$ and $\delta$. While $\epsilon$ is an upper bound on the degree of decoupling and depends only on the channel $\cF$, $\delta$ can be chosen arbitrarily small. One may hence think that the limit $\delta \rightarrow 0$ should be taken. This is true if the recovery error is the only concern. However, there is a trade-off relation between the recovery error and the circuit complexity, which is characterized by the parameter $\delta$. In fact, Eqs.~\eqref{eq: condition for t GYK} and~\eqref{eq: comp gYK} show that the circuit complexity of the generalized YK decoder depends on $\delta$, such as $\log (1/\delta)$. Hence, the complexity increases if one wishes to achieve small errors. This trade-off is naturally expected due to the implementation using quantum algorithms. Exponentially small $\delta$ is feasible since the dependence of the complexity on $1/\delta$ is only logarithmic.

Since $\cF^{AB \rarr D} = \cN^{C \rarr D}\circ \cE^{AB \rarr C}$, Theorem~\ref{thm: the gYK} states that when the encoding map $\cE$ is appropriately chosen against a given noise $\cN$, or equivalently when the encoder $\cE$ is chosen to satisfy the decoupling condition with small error, then the generalized YK decoder achieves a small error in recovering quantum information. 
In particular, consider the case where $\cN = \cN_1^{\otimes N}$ and $d_B = 1$.
As explained in~\ref{sec: decoupling and Uhlmann}, if the achievable rate $\mathrm{R}$ is below the quantum capacity $Q(\cN_1)$, there exists a suitable encoder that satisfies the decoupling condition with $\epsilon$ vanishing in the limit of increasing number of channel uses.
Hence, by setting $\delta$ in Theorem~\ref{thm: the gYK} to the value which vanishes in the limit, such as $1/d_A$, the generalized YK decoder can be used to achieve a rate arbitrarily close to the quantum capacity $Q(\cN_1)$.
This also applies to the entanglement-assisted setting with unlimited entanglement, while the achievable rate in this case approximates the entanglement-assisted quantum capacity $Q_{\rm E}(\cN_1)$.
Note that the generalized YK decoder remains explicit for any finite value of $N$.

In~\ref{sec: comparison}, we provide an in-depth comparison of the complexity with that of the algorithmic implementation of the Petz recovery map. We here mention that the number $t$ is dominant unless $\cC(U_\cF)$ is exponentially large. The number $t$ arises from the QSVT-based FPAA algorithm and is known to be an optimal order \cite{aaronson2012mony, yoder2014fixed, Gilyn2019thesis}. Hence, the quantum circuit implementation of the generalized YK decoder cannot be significantly improved. Note that,
while $t$ is independent of the choice of the dilation of $\cF^{AB \rarr D}$, the whole complexity is dependent on the choice due to the factor $\cC(U_\cF) + \log(d_D^2d_E/d_B)$ in Eq.~\eqref{eq: comp gYK}. Hence, using the unitary $U_\cF^L$ which minimizes $\cC(U_\cF) + \log(d_D^2d_E/d_B)$ results in the smallest complexity.

Another factor to be noted in the complexity is $\sqrt{d_D/d_B}$ in Eq.~\eqref{eq: condition for t GYK}, where $d_D$ is the dimension of the output of the noisy channel $\cN^{C \rarr D}$ and $d_B$ is that of the pre-shared entanglement. 
In the simplest case, where the encoding map is given by a unitary on $AB$ that is set to the same size as the input system $C$ of the noisy channel $\cN^{C \rarr D}$, we have $\sqrt{d_D/d_B} = \sqrt{d_Ad_D/d_C}$.
In this case, the complexity depends on $d_A$ and the ratio $d_D/d_C$ between the dimensions of the input $C$ and the output $D$ of the noisy channel. 
If the encoding is non-unitary, this is not the case, and one may expect that the complexity could be decreased by increasing $d_B$. This might be done by, e.g., fictitiously adding more entanglement at the outset, and by discarding it in the encoding process. This trick, however, does not affect the total complexity due to the other factor $[\lambda_{\rm min}(\omega^{RE})]^{-1/2}$. As $\ket{\omega}^{REDB'}$ is pure, $\lambda_{\rm min}(\omega^{RE}) = \lambda_{\rm min}(\omega^{DB'})$, where $\lambda_{\rm min}(\omega^{DB'})$ is non-zero minimum eigenvalue of $\omega^{DB'}$. This implies that, even if we fictitiously add extra entanglement of dimension $d_{\rm extra}$ for increasing $d_B$, the value of $\lambda_{\rm min}(\omega^{DB'})$ changes by factor $1/d_{\rm extra}$, which cancels the increase of $d_B$ in the complexity.\\


Before we move on, we explain the construction of the QSVT-based FPAA unitary $G_{t, \phi}^{DD'E'R'H}$ (see also~\ref{sec: explain QSVT FPAA}). To this end, we introduce two projectors:
\begin{align}
\label{eq: projection 1 GYK}
     &\Pi_1^{D'E'R'}  \coloneqq (V_\cF^{A'B'\rarr E'D'})^* \notag \\
     &\hspace{3pc}
    (\mathbb{I}^{B'} \otimes \ket{\Phi}\bra{\Phi}^{A'R'})(V_\cF^{A'B' \rarr E'D'})^\mathsf{T}, \\
     &\Pi_2^{DD'} \coloneqq \ket{\Phi}\bra{\Phi}^{DD'}, \label{eq: projection 2 GYK}
\end{align}
and unitaries:
\begin{equation}\begin{split}
\label{eq: rotation unitary W}
    W_m(\theta) 
    &\coloneqq e^{i\theta(2\Pi_m -\mathbb{I})},  \\
\end{split}\end{equation}
where $m = 1, 2$ and $\theta \in (-\pi, \pi]$.
Let $W_{t, \phi}^{DD'E'R'}$ be a unitary given by 
\begin{align}
\label{eq: prod W}
    &W_{t, \phi}^{DD'E'R'}  \coloneqq W_2(\phi_t)^{DD'}\prod_{j = 1}^{(t-1)/2}\Big[W_1(\phi_{2j})^{D'E'R'} \notag\\
    &\hspace{7pc}W_2(\phi_{2j-1})^{DD'}\Big].
\end{align}
The unitary $G^{DD'E'R'H}_{t, \phi}$ is then defined by
\begin{align}
\label{eq: FPAA unitary}
    &G_{t, \phi}^{DD'E'R'H}\coloneqq W_{t, \phi}^{DD'E'R'}\otimes \ketbra{+}{+}^H \notag\\
    &\hspace{6pc}+ W_{t, -\phi}^{DD'E'R'} \otimes \ketbra{-}{-}^H,
\end{align}
where $H$ is a single-qubit auxiliary system.
The unitary $G_{t, \phi}^{DD'E'R'H}$ has the block matrix representation as 
\begin{align}
\label{eq: after QSVT explain}
&G_{t, \phi}^{DD'E'R'H} \notag\\
&= 
\begin{pmatrix}
    Q_{t, \phi}(\Pi_2^{DD'}\Pi_1^{D'E'R'}) \otimes \ketbra{0}{0}^H & \cdot\hspace*{2mm} \\
    \cdot & \cdot \hspace*{2mm}
\end{pmatrix},
\end{align}
where $Q_{t, \phi}$ is a polynomial determined by degree $t$ and the phase sequence $\phi = (\phi_1, \ldots, \phi_t)$. 
When we choose an appropriate $t$ and $\phi$, the polynomial $Q_{t, \phi}$ well approximates the sign function, by which we can realize the QSVT-based FPAA.
More detailed analysis for its circuit complexity is explained in~\ref{sec: proof of det GYK}.


\subsection{Petz-like decoder}
\label{sec: IPM decoder}

Using a similar technique, we can construct the Petz-like decoder, which is a simplification of the Petz recovery map. 
We first introduce a decoding protocol with post-selection in \ref{sec: prob IPM}. Combining it with the QSVT-based FPAA algorithm, we explicitly construct the Petz-like decoder in \ref{sec: det IPM}~\footnote{While we assume that the state on $BB'$ is the MES $\ket{\Phi}^{BB'}$, the Petz-like decoder works even for an arbitrary state $\ket{\rho}^{BB'}$ (one example: the thermofield double state). In such cases, replacing every $\ket{\Phi}^{BB'}$ which appears in this section with $\ket{\rho}^{BB'}$ should suffice.}.


\begin{figure}
    \centering
    \includegraphics[width=80mm]{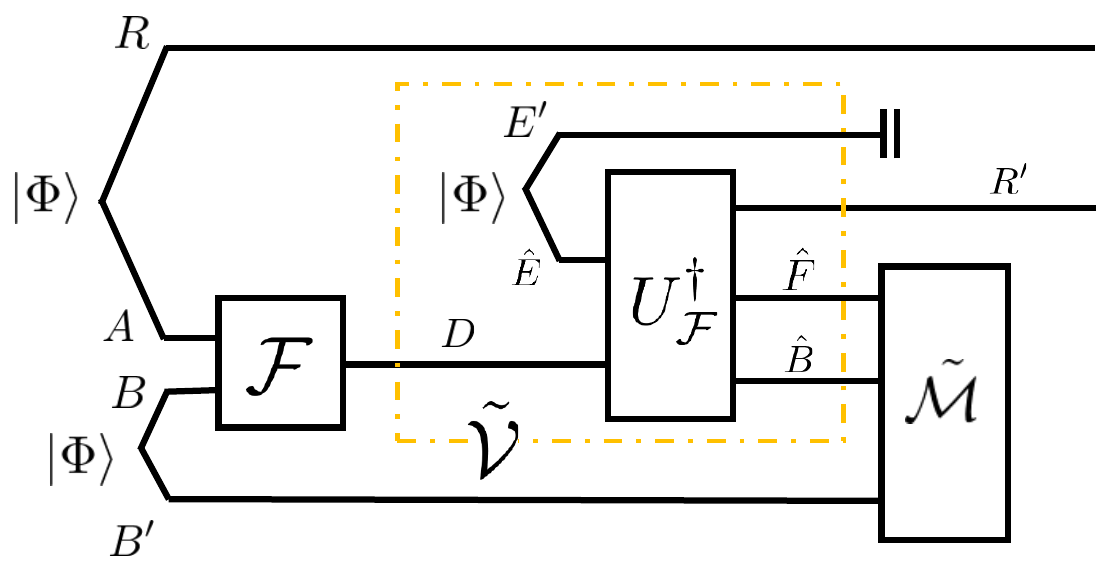}
    \caption{A diagram of the protocol with post-selection for the Petz-like decoder. The dash-dotted box represents the isometry map $\tilde{\cV}$ in Eq.~\eqref{eq: like post map}.}
    \label{fig: prob IPM}
\end{figure}

\subsubsection{Decoding protocol with post-selection}
\label{sec: prob IPM}

The decoding protocol with post-selection is as follows. 
See also Fig.~\ref{fig: prob IPM}.
Similarly to the generalized YK decoder, we denote a Stinespring isometry of $\cF^{AB \rarr D}$ by $U_\cF^L\ket{0}^F$ as given in Eqs.~\eqref{eq: stinespring iso-uni} and~\eqref{eq: stinespring unitary of F}.
Note that the protocol works for any choice of $U_\cF$.

\begin{enumerate}
    \item The receiver prepares ancilla qubits in the system $\hat{E}E'$, and then generates a MES $\Phi^{\hat{E}E'}$.
    \item The receiver applies the unitary $(U_\cF^{\hat{L}})^\dag$, where $\hat{L} = R'\hat{F}\hat{B} = \hat{E}D$.
    \item The receiver performs a binary measurement $\til{\cM} \coloneqq \{ \ketbra{0}{0}^{\hat{F}} \otimes \ketbra{\Phi}{\Phi}^{\hat{B}B'},  \mathbb{I}^{\hat{F}\hat{B}B'}   - \ketbra{0}{0}^{\hat{F}} \otimes \ketbra{\Phi}{\Phi}^{\hat{B}B'}\}$ on $\hat{F}\hat{B}B'$. When the former result of the measurement $\til{\cM}$ is obtained, this protocol succeeds.
\end{enumerate}
In this protocol, all the systems with a prime or a hat, and the channel output $D$, are in the hands of the receiver.
Below, we denote by $\til{\cV}^{D \rarr E'R'\hat{F}\hat{B}}$ an isometry map of the operation up to the step 2; that is, 
\begin{equation}\begin{split}
\label{eq: like post map}
    &\til{\cV}^{D \rarr E'R'\hat{F}\hat{B}}(\cdot)\coloneqq (U_\cF^{\hat{L}})^{\dag}(\cdot \otimes  \Phi^{\hat{E}E'})U_\cF^{\hat{L}}.
\end{split}\end{equation}

Conditioned by the success of the measurement $\til{\cM}$, the reduced state on the system $RR'$ is given by
\begin{align}
\label{eq: Plike post state}
    \til{\zeta}_{\rm succ}^{RR'} &= \tr_{E'\hat{E}\hat{B}B'}\Big[\f{1}{\til{p}_{\rm succ}}(\ketbra{0}{0}^{\hat{F}} \otimes \ketbra{\Phi}{\Phi}^{\hat{B}B'} ) \notag\\
    &\hspace{6pc}\til{\cV}^{D \rarr E'R'\hat{F}\hat{B}}(\omega^{RDB'})\Big],
\end{align}
where $\til{p}_{\rm succ}$ is the success probability of $\til{\cM}$, and $\omega^{RDB'} = \cF^{AB\rarr D}(\Phi^{AR}\otimes\Phi^{BB'})$.
It is straightforward to show that
\begin{align}
\label{eq: post select P like}
    &\til{p}_{\rm succ} = \f{d_A}{d_E}2^{-{H}_2(RE)_{\omega}}, \\
\label{eq: post fidelity plike}
    &\mathrm{F}\big(\til{\zeta}_{\rm succ}^{RR'}, \Phi^{RR'}\big) = \f{1}{d_A}2^{H_2(RE)_\omega - H_2(E)_\omega}.
\end{align}
See~\ref{sec: proof of Plike} for the details.

As mentioned before, $U_\cF^L$ is not uniquely determined from $\cF^{AB \rarr D}$.
Although this decoding protocol works for any choice of $U_\cF$, the decoding performance depends on the choice, which is unlike the generalized YK decoder. 
In fact, the success probability $\til{p}_{\rm succ}$ is inverse-proportional to $d_E$, which implies that it succeeds with higher probability if a smaller environment of the channel $\cF^{AB \rarr D}$ is chosen. Even though decoupling is satisfied, the probability $\til{p}_{\rm succ}$ is exponentially small.
On the other hand, the fidelity is the same as that of the generalized YK decoder. It is independent of the choice of $U_\cF$, and we have $\mathrm{F}\big(\til{\zeta}_{\rm succ}^{RR'}, \Phi^{RR'}\big) \approx 1$ when the decoupling is satisfied as $\omega^{RE} \approx \pi^R \otimes \omega^E$.


\subsubsection{Construction of the Petz-like decoder}
\label{sec: det IPM}

We now use the QSVT-based FPAA algorithm to amplify the success probability of the measurement $\til{\cM}$. For the same reasons as the generalized YK decoder, the amplification cannot be achieved with other known AA-type algorithms.

To describe the unitary $\til{G}_{t, \phi}$  corresponding to the QSVT-FPAA, let us define two projectors as
\begin{align}
\label{eq: two projection of Plike}
    &\til{\Pi}_1^{E'R'\hat{F}\hat{B}} 
    \coloneqq (U_\cF^{\hat{L}})^\dag (\ketbra{\Phi}{\Phi}^{\hat{E}E'}  \otimes \mathbb{I}^{D})U_\cF^{\hat{L}}, \\
\label{eq: prod 2 plike}
    &\til{\Pi}_2^{\hat{F}\hat{B}B'} \coloneqq \ketbra{0}{0}^{\hat{F}} \otimes \ketbra{\Phi}{\Phi}^{\hat{B}B'}.
\end{align}
By replacing $\Pi_m$ with $\til{\Pi}_m$ in the definition of $W_m(\theta)$ ($m=1,2$) in Eq.~\eqref{eq: rotation unitary W} and the following the constructions by Eqs.~\eqref{eq: prod W} and \eqref{eq: FPAA unitary}, we define the unitary $\til{G}_{t, \phi}^{E'R'\hat{F}\hat{B}B'H}$.

\begin{figure}
    \centering
    \includegraphics[width= 80mm]{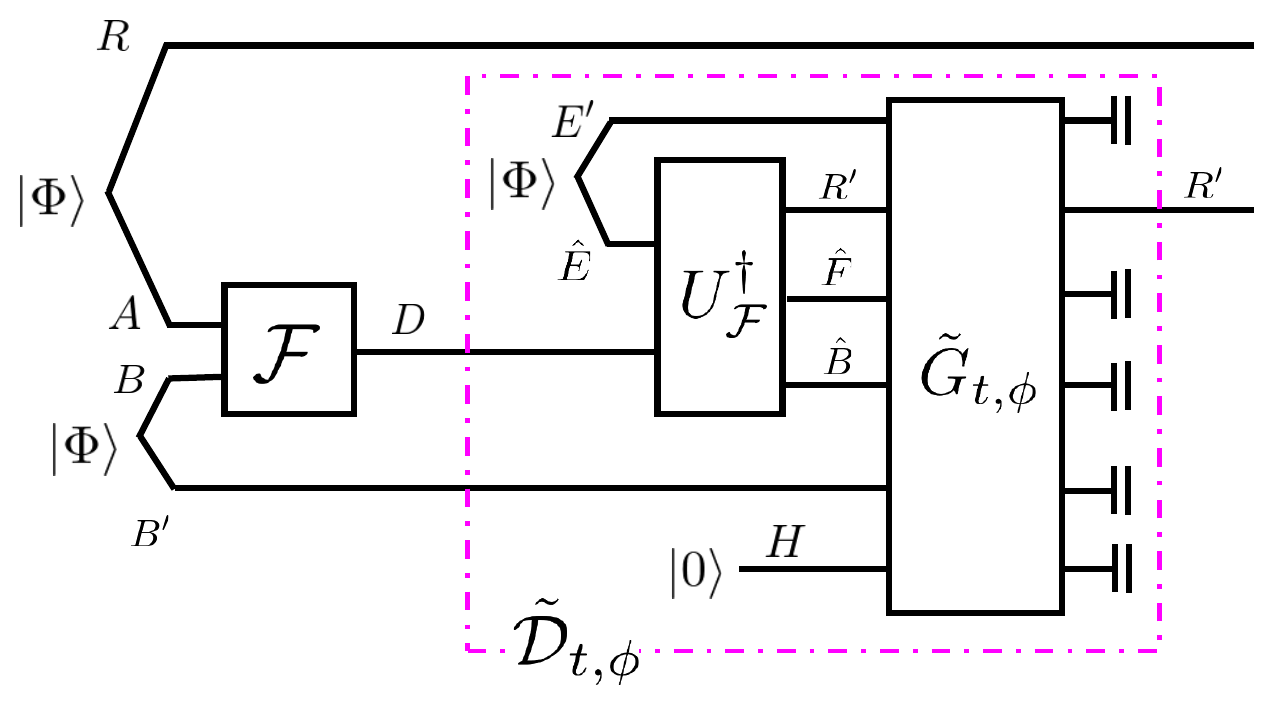}
    \caption{A diagram of the Petz-like decoder $\tilde{\cD}_{t, \phi}$, which is given in Eq.~\eqref{eq: def. plike decoder}, corresponds to the dash-dotted box. Note that $\til{G}_{t, \phi}$ consists of repeated applications of unitaries, which is similar to Fig.~\ref{fig: det GYK}.}
    \label{fig: det IPM decoder}
\end{figure}

The Petz-like decoder $\til{\cD}_{t, \phi}^{DB'\rarr R'}$ is given by replacing the step~3 with the following. See Fig.~\ref{fig: det IPM decoder} as well.
\begin{enumerate}
    \item[3'.] The receiver prepares an auxiliary state $\ket{0}^H$ in the system $H$ and applies the unitary $\til{G}_{t, \phi}^{E'R'\hat{F}\hat{B}B'H}$.
\end{enumerate}
With this modification, the Petz-like decoder is explicitly given by
\begin{align}
\label{eq: def. plike decoder}
    &\til{\cD}_{t, \phi}^{DB'\rarr R'}(\cdot) \notag
  \\
  &\hspace{0.5pc}\coloneqq \tr_{E'\hat{F}\hat{B}B'H}\big[\tilde{G}_{t, \phi}^{E'R'\hat{F}\hat{B}B'H} \big(\til{\cV}^{D \rarr E'R'\hat{F}\hat{B}}(\cdot)  \notag\\
  &\hspace{5pc}\otimes\ketbra{0}{0}^H\big)(\til{G}_{t, \phi}^{E'R'\hat{F}\hat{B}B'H})^\dag\big].
\end{align}
 The number $t \in \mathbb{N}$ and the phases $\phi \in (-\pi, \pi]^{t}$ are chosen such that the QSVT realizes an approximation of the sign function. 

The following theorem provides the performance of the Petz-like decoder.

\begin{theorem}[Performance of the Petz-like decoder]
\label{thm: det IPM}
For a given channel $\cF^{AB \rarr D}$, let $\bar{\cF}^{AB \rarr E}$ be its complementary channel, $\omega^{RE}$ be the state given by
\begin{equation}
    \omega^{RE} = \bar{\cF}^{AB \rarr E}(\Phi^{AR} \otimes \pi^B),
\end{equation}
and $\lambda_{\rm min}(\omega^{RE})$ be the non-zero minimum eigenvalue of $\omega^{RE}$.
Suppose that there exists a state $\tau^E$ such that $\|\omega^{RE} - \pi^R \otimes \tau^E\|_1 \leq \epsilon$. Then, for any $\delta \in (0, 1)$, and any odd integer $t$ satisfying
\begin{equation}
\label{eq: condition for t IPM}
    t \geq \bigg\lceil 16e\sqrt{\f{d_E}{d_A\lambda_{\rm min}(\omega^{RE})}} \log(2/\delta)\bigg\rceil,
\end{equation}
there exist $\phi = (\phi_1, \phi_2, \ldots, \phi_t) \in (-\pi, \pi]^t$ such that 
the recovery error $\Delta(\til{\cD}_{t, \phi}|\cF)$ of the Petz-like decoder $\til{\cD}_{t, \phi}^{DB' \rarr R'}$ is given by
    \begin{equation}
\label{eq: recovery error of Plike}
        \Delta(\til{\cD}_{t, \phi}|\cF) \leq \sqrt{\epsilon} + \delta,
    \end{equation}
and the circuit complexity of $\til{\cD}_{t, \phi}^{DB' \rarr R'}$ is 
\begin{equation}
\label{eq: comp P like}
    \cC\big(\til{\cD}_{t, \phi}\big) = \cO\Big( t\big(\cC(U_\cF) + \log (d_Dd_E^2/d_A)\big)\Big),
\end{equation}
and $\cO\big(\log(d_Dd_E^2/d_A)\big)$ ancilla qubits suffice.
Here, $\cC(U_\cF)$ is circuit complexity of a unitary $U_\cF^L$ such that $U_\cF^L\ket{0}^F$ is 
the Stinespring isometry of $\cF^{AB \rarr D}$, and $L = ABF = ED$.
\end{theorem}

\renewcommand{\arraystretch}{1.9}
\begin{table*}
\centering
\caption{A table of notation that we use in \ref{sec: comparison}. We use the numbers of qubits in the systems.}
\begin{tabular}{wc{3em}|wl {37em}} 
    $k$ & \hspace{3.2mm} The number of logical qubits in $A$: $k = \log d_A$. \\ \hline
    $n_{\rm in}$ & \hspace{3.2mm} The number of input qubits of the channel $\cN$: $n_{\rm in} = \log d_C$. \\ \hline
    $n_{\rm out}$ & \hspace{3.2mm} The number of output qubits of the channel $\cN$: $n_{\rm out} = \log d_D$. \\ \hline
    $b$ & \hspace{3.2mm} The number of ebits shared by the sender and the receiver in advance: $b = \log d_B$. \\ \hline
    $\kappa$ & 
    \renewcommand{\arraystretch}{1.4}
    \begin{tabular}{l}
    \hspace{0mm} The number of qubits in the environment $E$,\\ \hspace{0.1mm} which is equal to the logarithm of \#Kraus ops.: $\kappa = \log d_E = \log (\text{\#Kraus ops.})$. 
\end{tabular}
\renewcommand{\arraystretch}{1.9}
\end{tabular}
\label{fig: qubits number}
\end{table*}
\renewcommand{\arraystretch}{1.0}

Theorem~\ref{thm: det IPM} has many similarities to Theorem~\ref{thm: the gYK} about the generalized YK decoder, such as that the recovery error depends on the degree $\epsilon$ of decoupling as well as the parameter $\delta$ that characterizes the trade-off relation between the recovery error and the circuit complexity of the decoder.
From the upper bound on the recovery error in Eq.~\eqref{eq: recovery error of Plike}, we observe that, with a suitable encoder and $\delta$, the Petz-like decoder can also be used to achieve a rate arbitrarily close to the quantum capacity by increasing the number of channel uses, in the entanglement-non-assisted setting. 
In the entanglement-assisted setting with unlimited entanglement, it achieves a rate arbitrarily close to the entanglement-assisted quantum capacity.

On the other hand, the complexity of the Petz-like decoder differs from that of the generalized YK decoder. The number $t$, as well as the remaining part in $\cC(\til{\cD}_{t, \phi})$, explicitly depends on $d_E$. This implies that the complexity depends on the choice of the dilation of $\cF^{AB \rarr D}$, which reflects the aforementioned fact that the success probability of the protocol with post-selection is dependent on $d_E$.
Hence, it is desirable to use a dilated unitary $U_\cF^L$ with a small environment $E$. 
In the next section, we compare in detail the complexities of decoders and clarify in what cases one decoder has smaller complexity than the other.

As we will explain in the next section, the Petz-like decoder has a smaller circuit complexity than the algorithmic implementation of the original Petz recovery map \cite{gilyen2022petzmap}, when $\delta$ is appropriately chosen.
This is for two reasons. First, the Petz-like decoder is not exactly the same as the Petz recovery map. Although the Petz recovery map is known to be a good decoder, it is not necessary to implement the full map if one is interested in using the map as a decoder. This is one of the implications of our results. Second, the algorithmic implementation of the Petz recovery map \cite{gilyen2022petzmap} relies on the direct use of the QSVT to implement three CP maps that compose the Petz recovery map, which leads to high complexity. 
Due to the aforementioned simplification, we can cleverly use the QSVT-based FPAA instead of such direct uses of the QSVT, resulting in smaller computational cost.


\subsection{Comparison of the circuit complexities}
\label{sec: comparison}

We compare the circuit complexities of the generalized YK decoder, the Petz-like decoder, and the algorithmic implementation of the original Petz recovery map~\cite{gilyen2022petzmap}.
We derive a simple criterion that ensures the generalized YK decoder has smaller complexity compared to the Petz-like decoder.
We also demonstrate that the Petz-like decoder has substantially smaller complexity than the algorithmic implementation of the original Petz recovery map~\cite{gilyen2022petzmap}.

In the comparison, we sometimes use the number of qubits in each system instead of the dimensions.
We denote the number of qubits in $A$, $B$, $C$, $D$, and $E$ by $k$, $b$, $n_{\rm in}$, $n_{\rm out}$, and $\kappa$, respectively. See Table~\ref{fig: qubits number}.
Note that $\kappa$ is the logarithm of the number of the Kraus operators of the channel $\cF^{AB \rightarrow D}$, i.e., $\kappa = \log d_E = \log(\#\text{Kraus ops.})$. While this number depends on how the channel is dilated, we take the minimum possible number of Kraus operators in the comparison below, as we are interested in minimizing the complexity.

We first compare the complexity of the generalized YK decoder with that of the Petz-like decoder.
As explained in~\ref{sec: det GYK}, the number $t$ is the significant factor in the complexity.
We denote the numbers $t$ for the generalized YK decoder and for the Petz-like decoder by $t_{ \rm gYK}$ and $t_{ \rm Pl}$, respectively. That is,
\begin{align}
    &t_{ \rm gYK} = \Theta\Big(\big[2^{b - n_{\rm out}}\lambda_{\rm min}(\omega^{RE})\big]^{-1/2} \log(1/\delta)\Big),   \\
    &t_{ \rm Pl} = \Theta\Big(\big[2^{k - \kappa}\lambda_{\rm min}(\omega^{RE})\big]^{-1/2}  \log(1/\delta)\Big).
\end{align}
See Eq.~\eqref{eq: condition for t GYK} and Eq.~\eqref{eq: condition for t IPM}. 
Comparing $t_{ \rm gYK}$ and $t_{ \rm Pl}$, we find that
\begin{align}
\label{eq: boundary of comp}
   t_{\rm gYK} \leq t_{\rm Pl} 
   \iff k - b \leq \kappa - n_{\rm out}.
\end{align}
The left-hand side of the right condition in Eq.~\eqref{eq: boundary of comp} is given by the number of logical qubits $k$ that the sender intends to transmit and the number of pre-shared ebits $b$. On the other hand, the right-hand side depends on the quantities $\kappa$ and $n_{\rm out}$ that are the properties of the channel $\cF^{AB \rarr D}$. To better understand the condition in Eq.~\eqref{eq: boundary of comp}, we consider a couple of concrete instances below, assuming an isometric encoder for simplicity. In these cases, $\kappa$ corresponds to the logarithm of the number of Kraus operators of the noisy channel~$\cN^{C \rarr D}$.

For a given noisy channel $\cN^{C \rarr D}$, the right-hand side of Eq.~\eqref{eq: boundary of comp} represents a property of the noise. Hence, the number of logical qubits $k$ and that of pre-shared entanglement $b$ determine which decoder has smaller complexity. 
In general, the generalized YK decoder has an advantage when $b$ is large, and as $b$ becomes smaller, the advantage shifts to the Petz-like decoder. 
To observe this more concretely, we note that $0 \leq b \leq n_{\rm in} - k$. 
When the sender and the receiver pre-share the maximal amount of entanglement, i.e., $b = n_{\rm in}- k$, Eq.~\eqref{eq: boundary of comp} is rephrased as $k \leq \f{1}{2}(n_{\rm in} - n_{\rm out} + \kappa)$. In particular, if the input and the output systems of the channel $\cN^{C \rarr D}$ are identical, i.e., $n_{\rm in} = n_{\rm out}$, it reduces to 
\begin{equation}
    t_{\rm gYK} \leq t_{\rm Pl} \iff k \leq \f{1}{2}\kappa.    
\end{equation}
Hence, the generalized YK decoder has smaller complexity than the Petz-like decoder unless the number of logical qubits exceeds half of the logarithm of the number of Kraus operators of the noisy channel. 

In contrast, when no entanglement is shared in advance: $b=0$, Eq.~\eqref{eq: boundary of comp} reduces to 
\begin{equation}
    t_{\rm gYK} \leq t_{\rm Pl} \iff k \leq \kappa -  n_{\rm out}.    
\end{equation} 
Whether the right-hand side holds or not depends on the details of the noise. For instance, the amplitude damping on each qubit violates the inequality on the right-hand side. For such noises or the choice of large $k$, the Petz-like decoder has smaller complexity than the generalized YK decoder.

We may also use the fact that $k$ should necessarily satisfy $k \leq n_{\rm in}$ for the recovery to be possible. This leads to a trivial inequality
$k + n_{\rm out} - \kappa \leq n_{\rm in} + n_{\rm out} - \kappa$.
Furthermore, $\kappa$ always satisfies $\kappa \leq n_{\rm in} + n_{\rm out}$, since $\kappa$ is the logarithm of the number of Kraus operators. If a given noisy channel $\cN$ has the property that $\kappa = n_{\rm in} + n_{\rm out}$, it follows that 
\begin{equation}
    k + n_{\rm out} - \kappa \leq 0 \leq b,    
\end{equation} 
for any $b$. Hence, for the noise with the maximum possible number of Kraus operators, the generalized YK decoder has smaller complexity than the Petz-like decoder no matter how much entanglement is pre-shared.\\

We next compare the complexity of the Petz-like decoder with an algorithmic implementation of the original Petz recovery map provided in~\cite{gilyen2022petzmap}.
The following corollary can be derived by applying the algorithmic implementation to the Petz recovery map in our setting. In this corollary, $\|\cdot\|_\diamond$ denotes the diamond norm~\cite{Kitaev1997QCAlgoEC, watrous2004notessuperoperatornorm, watrous2018TheoryQI}.

\begin{corollary}[Algorithmic implementation of the Petz recovery map \cite{gilyen2022petzmap}]
    Let $\cP_{\pi, \cG}^{DB'\rarr R'}$ be the decoder based on the Petz recovery map defined in Eq.~\eqref{Eq:PetzOurSetting}.
    Then, there exists a quantum algorithm realizing the map $\til{\cP}_{\pi, \cG}^{DB'\rarr R'}$, which satisfies 
    \begin{equation}
    \label{eq: petz diamond}
    \|\til{\cP}_{\pi, \cG}^{DB'\rarr R'} - \cP_{\pi, \cG}^{DB'\rarr R'}\|_{\diamond} \leq \varepsilon,
    \end{equation}
    with circuit complexity  
    \begin{align}
    \label{eq: comp petz}
        &\cC(\til{\cP}_{\pi, \cG}) \notag\\
        &= \cO\Big(t_{ {\rm Petz}}\Big(\cC(U_\cF) + \log{(d_Bd_E)} + \f{\cC(U_\omega)}{\lambda_{\rm min}(\omega^{RE})} \notag\\
        &\hspace{2pc}\log{\big(\f{d_E}{\varepsilon}\big)} + d_A\log{(d_A)}\log{\big(\f{d_E}{\varepsilon\lambda_{\rm min}(\omega^{RE})}\big)}\Big)\Big),
    \end{align}
    where $t_{{\rm Petz}}$ is given by 
    \begin{equation}
        t_{{\rm Petz}} = \Bigg\lfloor\pi\sqrt{\f{d_E}{\lambda_{\rm min}(\omega^{RE})}}\Bigg\rfloor,
    \end{equation}
    and $\cC(U_{\omega})$ is circuit complexity of a unitary $U_\omega^{DB'P}$ such that, for any system $P$,
    \begin{equation}
    \label{eq: U_omega condition}
        \omega^{DB'} = \tr_P[U_\omega^{DB'P}\ketbra{0}{0}^{DB'P}(U_\omega^{DB'P})^\dag].
    \end{equation}  
\end{corollary}

From Eqs.~\eqref{cor: recovery error of Petz} and~\eqref{eq: petz diamond}, the recovery error of $\til{\cP}_{\pi, \cG}$ is bounded as
\begin{equation}
    \Delta(\til{\cP}_{\pi, \cG}|\cF) \leq 2\epsilon^{1/4} + \varepsilon,
\end{equation}
when there exists $\tau^E$ such that $\|\omega^{RE} - \pi^R \otimes \tau^E\|_1 \leq \epsilon$.

We clarify the condition that the Petz-like decoder has smaller complexity than the algorithmic implementation of the Petz recovery map.
First, when $\cC(U_\cF)$ is larger than the other terms, Eqs.~\eqref{eq: comp P like} and~\eqref{eq: comp petz} approximately reduce to
\begin{align}
&\cC\big(\til{\cD}_{t, \phi}\big) \approx \cO\big(t_{\rm Pl}\cC(U_\cF)\big), \\
&\cC\big(\til{\cP}_{\pi, \cG}\big) \approx \cO\big(t_{\rm Petz}\cC(U_\cF)\big),
\end{align}
respectively.
In this case, we only need to compare $t_{\rm Pl}$ with $t_{ \rm Petz}$, which satisfies $t_{ \rm Pl} = \Theta\big(\f{\log(1/\delta)}{\sqrt{d_A}}t_{ \rm Petz}\big)$. Hence, as far as $\delta = \Omega( 2^{-\sqrt{d_A}})$, the Petz-like decoder has smaller complexity than the algorithmic implementation of the original Petz recovery map. 
For instance, by taking $\delta = 1/d_A$, the Petz-like decoder achieves a reduced circuit complexity by a leading factor of $\sqrt{d_A} = 2^{k/2}$, implying that the exponent of the exponential scaling in the circuit complexity changes to a smaller one.

\renewcommand{\arraystretch}{1.9}
\begin{table*}
\centering
\caption{The circuit complexity of our decoders in particular noise models. We denote ${\min}_i p_i$ by $p_{\rm min}$. The constant $\gamma$ is assumed to be $1/2$ or less. We have assumed a unitary encoding by a polynomial-sized quantum circuit, so $k + b = n_{\rm in}$. 
We have also assumed the decoupling condition $\omega^{RE} \approx \pi^R \otimes \omega^E$, which leads to $\lambda_{\rm min}(\omega^{RE}) \approx \lambda_{\rm min}(\omega^E)/d_A$ with $\lambda_{\rm min}(\omega^E)$ being the non-zero minimum eigenvalue of $\omega^{E} = \bar{\cN}^{C \rarr E}(\pi^C)$.
The part $\mathrm{poly}(\cdot)$ comes from the term associated with the dilated unitary of the noise, and from the term logarithmic in dimensions in Eqs.~\eqref{eq: comp gYK} and~\eqref{eq: comp P like}.}
\begin{tabular}{wc{11em}|wc{15em}|wc{15em}} 
                & Generalized YK decoder $\cC\big(\cD_{t, \phi}\big)$ & Petz-like decoder $\cC\big(\til{\cD}_{t, \phi}\big)$ \\ \hline
    Pauli noise &   $\big[\big(2^k/p_{\rm min}^{n/2}\big)\log(1/\delta)\big]\mathrm{poly}(n, k)$ & $\big[\big(2/p_{\rm min}^{1/2}\big)^n \log(1/\delta)\big]\mathrm{poly}(n, k)$ \\ \hline
    Amplitude damping noise & $\big[2^k(2/\gamma)^{n/2}\log(1/\delta)\big] \mathrm{poly}(n, k)$ & $\big[(4/\gamma)^{n/2} \log(1/\delta)\big]\mathrm{poly}(n, k)$ \\ \hline
    Erasure noise & $\big[2^k\log(1/\delta)\big]\mathrm{poly}(n_{\rm in}, n_{\rm out}, k)$ & $\big[2^{n_{\rm in} - n_{\rm out}}\log(1/\delta)\big]\mathrm{poly}(n_{\rm in}, n_{\rm out}, k)$ \\ 
\end{tabular}
\label{fig: specific noise complexity}
\end{table*}
\renewcommand{\arraystretch}{1.0}

The advantage of the Petz-like decoder remains even when $\cC(U_\cF)$ is not dominant. To see this, suppose that $\varepsilon$ in Eq.~\eqref{eq: comp petz} is $\varepsilon = \cO(\delta)$ with sufficiently small $\delta$. 
The complexity of the algorithmic implementation of the Petz recovery map reduces to
\begin{align}
    \cC\big(\til{\cP}_{\pi, \cG}\big) 
    &\approx \cO\biggl(t_{\rm Petz} \log(1/\delta)\mathrm{poly}(n_{\rm in}, n_{\rm out}, k) \notag \\
    &\hspace{4pc} \Big(\f{1}{\lambda_{\rm min}(\omega^{RE})} + 2^k\Bigr)   \biggr) \\
    &=\cO\biggl(t_{\rm Pl} \mathrm{poly}(n_{\rm in}, n_{\rm out}, k) \notag \\
    &\hspace{3pc} 2^{k/2}\Big(\f{1}{\lambda_{\rm min}(\omega^{RE})} + 2^k\Bigr)   \biggr).
    \label{eq: approx comp small delta petz}
\end{align}
Here, we used in the second equation that $t_{ \rm Pl} = \Theta\big(\f{\log(1/\delta)}{\sqrt{d_A}}t_{ \rm Petz}\big)$ and assumed that $\cC(U_\cF^L)$ is polynomial in qubits, which further implies that $\cC(U_\omega^{DB'P})$ is polynomial.
On the other hand, the complexity of the Petz-like decoder in this case is  
\begin{align}
\label{eq: approx comp small delta plike}
    \cC\big(\til{\cD}_{t, \phi}\big) = \cO\big(t_{ \rm Pl}\mathrm{poly}(n_{\rm in}, n_{\rm out}, k)\big).
\end{align}
Since this corresponds to the first line of Eq.~\eqref{eq: approx comp small delta petz}, the Petz-like decoder has smaller circuit complexity than the algorithmic implementation of the Petz recovery map.
As the second line of Eq.~\eqref{eq: approx comp small delta petz} is at least an order of $2^{k/2}$, a significant reduction in the circuit complexity is achieved in this case as well.

From these comparisons of explicit quantum circuit implementations, we conclude that,  for isometry encoding and noisy channels with the maximum number of Kraus operators, the circuit complexity increases in the following order: the generalized YK decoder, the Petz-like decoder, and the algorithmic implementation of the Petz recovery map. For channels with fewer Kraus operators, the Petz-like decoder may have smaller complexity than the generalized YK decoder, depending on more specific noise properties. However, both decoders have smaller complexity than the algorithmic implementation of the Petz recovery map.

Note that, although all the complexities are exponential in the number of qubits, this must be the case as it is in general computationally hard to decode quantum information~\cite{Vardy1997intractability, Kuo2012hardnessdepo, Iyer2015hardness}.


\subsection{Application to concrete noisy models}
\label{sec: specific compare}

We consider several noises for demonstration. We investigate the noises that independently act on each qubit, such as 
the independent Pauli noise, the independent amplitude damping noise, and the qubit-erasure noise. 
If the input system $C$ of the noisy channel $\cN^{C \rarr D}$ is equal to the output system $D$ of it, we denote by $S$ the system as $S = C = D$, and by $n$ the number of these qubits as $n = n_{\rm in} = n_{\rm out}$.  

In Table~\ref{fig: specific noise complexity}, we summarize the circuit complexities of our decoders. From these results, we find that the complexities become smaller in more noisy situations, such as for large $p_{\rm min} = \min_{i = 0, 1, 2, 3} p_i$ in the Pauli noise or large $\gamma$ in the amplitude damping noise.

\begin{itemize}
    \item Independent Pauli noise
\end{itemize}
The first example is the independent Pauli noise. A Stinespring isometry of the single-qubit Pauli noise is given by
\begin{equation}\begin{split}
    V_\cN^{S\rarr ES} = \sum_{i=0}^3 \sqrt{p_i} \ket{e_i}^{E} \otimes \sigma_i^S,
\end{split}\end{equation}
where $\sum_{i=0}^3p_i = 1$ and $(\sigma_i^S) = (\mathbb{I}^S, X^S, Y^S, Z^S)$. 
Since the number of qubits of the system $S$ is $n$, and the logarithm of the number of the Kraus operators $\kappa = 2n$, we can rephrase Eq.~\eqref{eq: boundary of comp} as
\begin{align}
\label{eq: pauli comp}
    &n -k +b \geq 0  \iff t_{\rm gYK} \leq t_{\rm Pl}.
\end{align}
Since $k \leq n - b$ is always satisfied, the generalized YK decoder has smaller complexity than the Petz-like decoder for the independent Pauli noise.

\begin{itemize}
    \item Independent amplitude damping noise
\end{itemize}
The second example is the amplitude damping noise for $\{\ket{0}^S, \ket{1}^S\}$, which independently acts on each qubit.
The single-qubit amplitude damping noise is represented by an isometry 
\begin{equation}\begin{split}
    &V_\cN^{S\rarr ES} = \sqrt{\gamma}\ket{e_0}^E \otimes \ketbra{0}{1}^S  \\
    &\hspace{4.5pc}+ \ket{e_1}^E \otimes\big(\ketbra{0}{0}^S + \sqrt{1-\gamma}\ketbra{1}{1}^S\big),        
\end{split}\end{equation}
where $\gamma \in [0, 1]$.
As $n = \kappa$, Eq.~\eqref{eq: boundary of comp} becomes 
\begin{align}
    b - k \geq 0  \iff t_{\rm  gYK} \leq t_{\rm Pl}.
\end{align}
Hence, when the number of pre-shared entanglement $b$ is more than the number of the logical qubits $k$, the generalized YK decoder has smaller complexity than the Petz-like decoder.

\begin{itemize}
    \item Qubit-erasure noise
\end{itemize}
The third example is the qubit-erasure noise, which erases $\kappa$ qubits out of $n_{\rm in}$ input qubits. The erased qubits are randomly chosen, but it is assumed that the receiver knows which qubits were erased.
In this case, it holds that $n_{\rm in} = n_{\rm out} + \kappa$. Thus, Eq.~\eqref{eq: boundary of comp} becomes  
\begin{align}
\label{eq: erasure comp}
    &n_{\rm in} - 2n_{\rm out} -k + b \geq 0 \iff t_{\rm  gYK} \leq t_{\rm Pl}.
\end{align}
Especially, when there is no pre-shared entanglement, $b = 0$, and the communication rate $k/n_{\rm in}$ is given by $k/n_{\rm in} = n_{\rm out}/n_{\rm in} - 1/2$, which is approximately half of the quantum capacity, Eq.~\eqref{eq: erasure comp} does not hold, and the Petz-like decoder has smaller complexity than the generalized YK decoder.
On the other hand, when the maximal amount of entanglement is pre-shared, i.e., $b = n_{\rm in} - k$, Eq.~\eqref{eq: erasure comp} is rephrased as $k \leq n_{\rm in} - n_{\rm out} = \kappa$. Hence, if more than $k$ qubits are erased by the noise, the generalized YK decoder has smaller complexity than the Petz-like decoder.

\section{Proofs}
\label{sec: proofs of these results}

In this section, we provide proofs of the main results. In \ref{sec: proof of GYK} and \ref{sec: proof of Plike}, we show the statements about the generalized YK decoder and the Petz-like decoder, respectively.

\subsection{Proofs: the generalized YK decoder}
\label{sec: proof of GYK}

We first consider the decoding protocol with post-selection, and provide the success probability and the fidelity after the post-selection. We then prove Theorem~\ref{thm: the gYK}.


\subsubsection{Success probability and fidelity in the decoding protocol with post-selection}
\label{sec: proof of prob GYK}

\begin{figure}
    \centering
    \includegraphics[width=80mm]{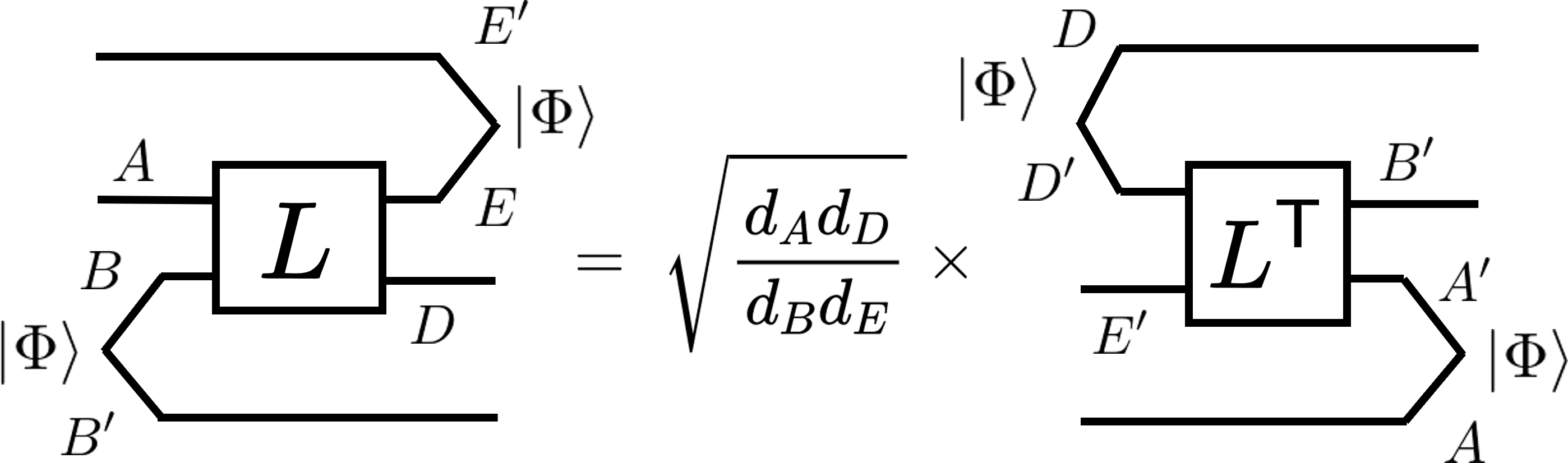}
    \caption{A diagram of the transpose of a matrix $L$ sandwiched by two MESs}
    \label{fig: transp MES}
\end{figure}

Before proving Theorem~\ref{thm: the gYK}, we first derive Eqs.~\eqref{eq: probability of gyk} and~\eqref{eq: post select GYK} here.
To this end, we introduce an important lemma, which will be used frequently in what follows.
Its proof is straightforward by direct calculation. See Fig.~\ref{fig: transp MES} for the diagram of the statement.

\begin{lemma}[Transpose of a matrix sandwiched by two MESs]
\label{lem: rotation lemma}
For any linear operator $L^{AB\rarr ED}$, i.e., $d_Ed_D\times d_Ad_B$ matrix, it holds that
    \begin{equation}\begin{split}
    \label{eq: trans matrix}
       &\bra{\Phi}^{EE'}\big(\mathbb{I}^{B'E'} \otimes L^{AB \rarr ED}\big)\ket{\Phi}^{BB'} \\
       &\hspace{0pc}=\sqrt{\f{d_Ad_D}{d_Bd_E}}\bra{\Phi}^{AA'}\big((L^{A'B' \rarr E'D'})^\mathsf{T}\otimes \mathbb{I}^{AD}\big)\ket{\Phi}^{DD'}.
    \end{split}\end{equation}
This is a linear operator from $AE'$ to $B'D$.
The transpose is taken with respect to the basis that defines each MES.
\end{lemma}

We consider our decoding protocol with post-selection.
The input state of the decoding protocol is given by 
\begin{equation}
\label{inteq:13}
    \omega^{RDB'} = \cF^{AB \rarr D}(\Phi^{AR}\otimes \Phi^{BB'}).
\end{equation}
By including the environment $E$, its purified state is
\begin{equation}
\label{inteq:12}
    \ket{\omega}^{REDB'}= V_\cF^{AB \rarr ED}\ket{\Phi}^{AR}\ket{\Phi}^{BB'}.
\end{equation}
Using Lemma \ref{lem: rotation lemma} for $L = V_\cF^*$, the state $\zeta_{\rm succ}^{RR'}$ on the system $RR'$ after the post-selection is rewritten as
\begin{align}
\zeta_{\rm succ}^{RR'} 
    &= \f{1}{p_{\rm succ}}\tr_{E'}\big[\bra{\Phi}^{DD'}(V_\cF^{A'B' \rarr E'D'})^* \notag \\
    &\hspace{1pc}(\omega^{RDB'}\otimes \Phi^{A'R'})(V_\cF^{A'B' \rarr E'D'})^{\mathsf{T}}\ket{\Phi}^{DD'}\big] \\
    &= \f{1}{p_{\rm succ}}\tr_{E'}\big[\bra{\Phi}^{DD'}(V_\cF^{A'B' \rarr E'D'})^*\ket{\Phi}^{A'R'} \notag \\
    &\hspace{2pc}\omega^{RDB'}\bra{\Phi}^{A'R'}(V_\cF^{A'B' \rarr E'D'})^{\mathsf{T}}\ket{\Phi}^{DD'}\big] \\
    &= \f{1}{p_{\rm succ}}\f{d_Bd_E}{d_Ad_D}\tr_{E'}\big[\bra{\Phi}^{\hat{B}B'} (V_\cF^{R'\hat{B} \rarr \hat{E}D})^\dag  \notag \\
    &\hspace{1pc}\ket{\Phi}^{\hat{E}E'}\omega^{RDB'}\bra{\Phi}^{\hat{E}E'}V_\cF^{R'\hat{B}\rarr \hat{E}D}\ket{\Phi}^{\hat{B}B'}\big] \\
    &= \f{1}{p_{\rm succ}}\f{d_B}{d_Ad_D}\bra{\Phi}^{\hat{B}B'} (V_\cF^{R'\hat{B} \rarr \hat{E}D})^\dag  \notag \\
    &\hspace{2pc}(\omega^{RDB'}\otimes \mathbb{I}^{\hat{E}})V_\cF^{R'\hat{B}\rarr \hat{E}D}\ket{\Phi}^{\hat{B}B'}.
\end{align}
Here, we used Lemma~\ref{lem: rotation lemma} in the third equation.
The success probability of the measurement $\cM$ is then given as 
\begin{align}
    p_{\rm succ} 
    &=\f{d_B}{d_Ad_D} \tr[\bra{\Phi}^{\hat{B}B'}\big(V_\cF^{R'\hat{B}\rarr \hat{E}D}\big)^\dag \notag \\
    &\hspace{2pc}(\omega^{RDB'} \otimes \mathbb{I}^{\hat{E}})V_\cF^{R'\hat{B}\rarr\hat{E}D}\ket{\Phi}^{\hat{B}B'}] \\
    &=\f{d_B}{d_D}\tr\big[\big(\mathbb{I}^R\otimes V_\cF^{R'\hat{B}\rarr\hat{E}D} (\pi^{R'} \otimes \Phi^{\hat{B}B'}) \notag \\
    &\hspace{3pc}(V_\cF^{R'\hat{B}\rarr\hat{E}D})^\dag\big)(\omega^{RDB'}\otimes \mathbb{I}^{\hat{E}})\big] \\
    &=\f{d_B}{d_D}\tr[(\mathbb{I}^R\otimes \omega^{DB'\hat{E}})(\omega^{RDB'}\otimes \mathbb{I}^{\hat{E}})] \\
    &=\f{d_B}{d_D}\tr[(\omega^{DB'})^2] \\
    &=\f{d_B}{d_D}2^{-H_2(RE)_\omega}. \label{eq: p rep.}
\end{align}
Since the state $\ket{\omega}^{REDB'}$ is pure, we here used that $\tr[(\omega^{DB'})^2] = \tr[(\omega^{RE})^2] = 2^{-H_2(RE)_{\omega}}$.

The fidelity after the post-selection is calculated from $\zeta^{RR'}_{\rm succ}$ as follows: 
\begin{align}
    &\mathrm{F}(\zeta_{\rm succ}^{RR'}, \Phi^{RR'}) \notag \\
    &= \f{1}{p_{\rm succ}} \f{d_B}{d_Ad_D} \tr[\Phi^{RR'}\bra{\Phi}^{\hat{B}B'} \big(V_\cF^{R'\hat{B}\rarr\hat{E}D}\big)^\dag \notag \\
    &\hspace{4pc}(\omega^{RDB'}\otimes \mathbb{I}^{\hat{E}})V_\cF^{R'\hat{B}\rarr\hat{E}D}\ket{\Phi}^{\hat{B}B'}] \\
    &= \f{1}{p_{\rm succ}} \f{d_B}{d_Ad_D} \tr[V_\cF^{R'\hat{B}\rarr\hat{E}D}(\Phi^{RR'}\otimes \Phi^{\hat{B}B'}) \notag \\
    &\hspace{5pc}(V_\cF^{R'\hat{B}\rarr\hat{E}D})^\dag(\omega^{RDB'}\otimes \mathbb{I}^{\hat{E}})] \\
    &=\f{1}{p_{\rm succ}} \f{d_B}{d_Ad_D}\tr[\omega^{R\hat{E}DB'}(\omega^{RDB'} \otimes\mathbb{I}^{\hat{E}})] \\
    &=\f{1}{p_{\rm succ}} \f{d_B}{d_Ad_D}\tr[(\omega^{RDB'})^2] \\
    &=\f{1}{p_{\rm succ}} \f{d_B}{d_Ad_D} 2^{-H_2(RDB')_\omega}.
\end{align}
Substituting Eq.~\eqref{eq: p rep.}, we obtain that
\begin{align}
   \mathrm{F}(\zeta_{\rm succ}^{RR'}, \Phi^{RR'}) 
   &=\f{1}{d_A}2^{H_2(RE)-H_2(RDB')} \\
   &=\f{1}{d_A}2^{H_2(RE)-H_2(E)},
\end{align}
where we used $H_2(RDB')_\omega = H_2(E)_\omega$ since $\ket{\omega}^{REDB'}$ is pure.
Thus, we obtain Eqs.~\eqref{eq: probability of gyk} and~\eqref{eq: post select GYK}.


\subsubsection{Proof of Theorem \ref{thm: the gYK}}
\label{sec: proof of det GYK}

To show Theorem \ref{thm: the gYK}, we use the QSVT-based FPAA algorithm instead of the measurement $\cM$.
We again mention that our situation differs from the common situation for the AA algorithm since the receiver has access only to a part of the whole system; the reference $R$ and environment $E$ are not with the receiver. This issue will be circumvented by the QSVT-based FPAA.

First, we specify the input and output states of the QSVT-based FPAA.
We denote the input state by
\begin{align}
    \omega_0^{RDD'E'R'} \coloneqq \cV^{B'\rarr D'E'R'}(\omega^{RDB'}),
\end{align}
where $\omega^{RDB'}$ is given by Eq.~\eqref{inteq:13}, and $\cV^{B'\rarr D'E'R'}$ is the isometry map such that 
\begin{equation}\begin{split}
    &\cV^{B'\rarr D'E'R'}(\cdot) \\
    &= (V_\cF^{A'B' \rarr E'D'})^*(\cdot\otimes \Phi^{A'R'})(V_\cF^{A'B' \rarr E'D'})^{\mathsf{T}}.
\end{split}\end{equation}
To introduce a target state, we define a pure state $\ket{\omega_{\rm targ}}^{RR'EE'}$ by 
\begin{align}
\label{inteq:8}
        \ket{\omega_{\rm targ}}^{RR'EE'}
        &\coloneqq \sqrt{d_Ad_E \omega^{RE}}\ket{\Phi}^{RER'E'}.
\end{align}
A reduced state on $RR'$, i.e., $\omega_{\rm targ}^{RR'}$, is our target state, to which we aim to transform the input state $\omega_0^{RDD'E'R'}$. This is because,
if there exists a state $\tau^E$ such that $\|\omega^{RE} - \pi^R \otimes \tau^E \|_1 \leq \epsilon$, it holds that 
\begin{equation}
\label{eq: target state of lemma}
    \frac{1}{2}\big\|\omega_{\rm targ}^{RR'} - \Phi^{RR'}\big\|_1 \leq \sqrt{\epsilon}.
\end{equation}
This follows from the following proposition, which considers, in a slightly more general case, the so-called \emph{canonical purification}~\cite{Winter2004ExtrinsicIntrinsic}.

The canonical purification for a state $\rho^A$ is defined as $\ket{\rho_{\rm c}}^{AA'} \coloneqq \sqrt{d_A \rho^A} \ket{\Phi}^{AA'} = \sum_i \sqrt{p_i}\ket{e_i}^A\ket{e_i^*}^{A'}$, where $\sum_i p_i \ketbra{e_i}{e_i}^A$ is the eigenvalue decomposition of $\rho^A$. Note that $\sqrt{d_A}\ket{\Phi}^{AA'} = \sum_i \ket{i}^A\ket{i}^{A'} = \sum_i \ket{e_i}^A\ket{e_i^*}^{A'}$, where $\{\ket{i}\}_i$ is the computational basis. The canonical purification satisfies the following.

\begin{proposition}
\label{lem: vectorization lemma general}
Let $\ket{\rho_{\rm c}}^{AA'}$ and $\ket{\sigma_{\rm c}}^{AA'}$ be the canonical purifications of states $\rho^A$ and $\sigma^A$, respectively. Then, it holds that
\begin{align}
    \frac{1}{2}\big\|\ketbra{\rho_{\rm c}}{\rho_{\rm c}}^{AA'} - \ketbra{\sigma_{\rm c}}{\sigma_{\rm c}}^{AA'}\big\|_1 \leq \sqrt{\|\rho^A - \sigma^A\|_1}.
\end{align}
\end{proposition}

While this proposition may be derived readily from results in~\cite{Audenaert2014comparison}, we provide a proof for completeness.
\begin{proof}[Proof of Proposition~\ref{lem: vectorization lemma general}]

Let us start by evaluating the Euclidean norm of the difference between $\ket{\rho_{\rm c}}^{AA'}$ and $\ket{\sigma_{\rm c}}^{AA'}$:
\begin{align}
    &\big\|\ket{\rho_{\rm c}}^{AA'} - \ket{\sigma_{\rm c}}^{AA'}\big\| \notag\\
    &= \big\|\sqrt{d_A\rho^A}\ket{\Phi}^{AA'} - \sqrt{d_A\sigma^A}\ket{\Phi}^{AA'}\big\| \\
    &= \sqrt{\tr[\rho^A] + \tr[\sigma^A] - 2\tr\big[\sqrt{\rho^A}\sqrt{\sigma^A}\big]} \\
    &=\big\|\sqrt{\rho^A} - \sqrt{\sigma^A}\big\|_2 \\
    &\leq \sqrt{\|\rho^A - \sigma^A\|_1}.
\end{align}
In the last line, we used the well-known Powers--St\o rmer inequality~\cite{powers1970free, kittaneh1987inequalities}: for any positive semidefinite operators $L$ and $M$, $\|L^{1/2} - M^{1/2}\|_2 \leq \|L - M\|_1^{1/2}$ holds.
Then, since $\f{1}{2}\|\ketbra{v}{v} - \ketbra{w}{w}\|_1 \leq \|\ket{v} - \ket{w}\|$ for any pure states $\ket{v}$ and $\ket{w}$, we obtain
\begin{equation}
    \f{1}{2}\big\|\ketbra{\rho_{\rm c}}{\rho_{\rm c}}^{AA'} - \ketbra{\sigma_{\rm c}}{\sigma_{\rm c}}^{AA'}\big\|_1 \leq \sqrt{\|\rho^A - \sigma^A\|_1}.
\end{equation}

\end{proof}

By taking $\rho^A$ and $\sigma^A$ in Proposition~\ref{lem: vectorization lemma general} as $\omega^{RE}$ and $\pi^R \otimes \tau^E$, respectively, and using the contraction property of the trace norm under the partial trace over $EE'$, we obtain Eq.~\eqref{eq: target state of lemma}.
This implies that the state $\omega_{\rm targ}^{RR'}$ is $\sqrt{\epsilon}$-close to the MES $\Phi^{RR'}$ when the decoupling is satisfied with error $\epsilon$.
Hence, we aim to transform the input state to the target state: $\omega_0^{RDD'E'R'} \mapsto \omega_{\rm targ}^{RR'} \approx \Phi^{RR'}$, ensuring that the recovery is successful when $\omega^{RE} \approx \pi^R \otimes \tau^E$.\\

In the following, our goal is to show that the transformation from $\omega_0^{RDD'E'R'}$ to $\omega_{\rm targ}^{RR'}$ can be achieved with high accuracy by the QSVT-based FPAA algorithm.
Before we consider the QSVT-based FPAA in full detail, we first examine the structures of the input and output states to construct an operator to which the QSVT-based FPAA should be applied.

Let $\ket{\omega_0}^{REDD'E'R'}$ be the purified state of $\omega_0^{RDD'E'R'}$, i.e., 
\begin{align}
    \ket{\omega_0}^{REDD'E'R'} 
    = (V_\cF^{A'B' \rarr E'D'})^*\ket{\omega}^{REDB'}\ket{\Phi}^{A'R'}.
\end{align}
The state $\ket{\omega}^{REDB'}$ is given by Eq.~\eqref{inteq:12}.
Suppose $\ket{\omega_0}^{REDD'E'R'}$ has the Schmidt decomposition between $RE$ and $DD'E'R'$ such that
\begin{equation}
    \begin{split}
    \label{eq: schmidt omega 0}
    \ket{\omega_0}^{REDD'E'R'} = \sum_{\mu=1}^r \sqrt{\lambda_\mu}\ket{\eta_\mu}^{RE}\ket{\psi_\mu}^{DD'E'R'},
    \end{split}
\end{equation}
where $\{\ket{\psi_\mu}^{DD'E'R'}\}_\mu$ and $\{\ket{\eta_\mu}^{RE}\}_\mu$ are orthonormal bases, and $r$ is the Schmidt rank, i.e., the number of non-zero $\lambda_\mu$'s.
From Eq.~\eqref{inteq:8} and the fact that $\omega^{RE} = \omega_0^{RE}$, it follows that the Schmidt decomposition of $\ket{\omega_{\rm targ}}^{RR'EE'}$ is given by
\begin{align}
\label{eq: def omega targ}
    \ket{\omega_{\rm targ}}^{RR'EE'}
    = \sum_\mu \sqrt{\lambda_\mu}\ket{\eta_\mu}^{RE}\ket{\eta_\mu^*}^{E'R'}. 
\end{align}
Thus, if we can transform $\ket{\psi_\mu}^{DD'E'R'}$ to $\ket{\eta_\mu^*}^{E'R'}$ for all $\mu$, we can obtain $\omega_{\rm targ}^{RR'}$ on $RR'$.

To this end, we now turn to investigate the QSVT-based FPAA. 
Consider an operator $\Lambda^{DD'E'R'}$ defined by $\Lambda^{DD'E'R'} \coloneqq \Pi_2^{DD'} \Pi_1^{D'E'R'}$, which is the subject of the QSVT-based FPAA in our protocol.
Here, $\Pi_1^{D'E'R'}$ and $\Pi_2^{DD'}$ are the projectors such that
\begin{align}
    \label{eq: proj gyk}
        &\Pi_{1}^{D'E'R'} 
        \coloneqq (V_\cF^{A'B' \rarr E'D'})^* (\mathbb{I}^{B'} \otimes\ket{\Phi}\bra{\Phi}^{A' R'}) \notag\\
        &\hspace{7pc}(V_\cF^{A'B' \rarr E'D'})^\mathsf{T},  \\
    \label{inteq:4}
        &\Pi_2^{DD'}
        \coloneqq \ket{\Phi}\bra{\Phi}^{DD'}.
\end{align}
To apply the QSVT, we need to specify the singular value decomposition of $\Lambda^{DD'E'R'}$. This can be done as follows.

We first calculate $(\Pi_1\Pi_2\Pi_1)^{DD'E'R'}$: 
 \begin{align}
            &(\Pi_1\Pi_2\Pi_1)^{DD'E'R'} \notag \\
            &= (V_\cF^{A'B' \rarr E'D'})^*\ketbra{\Phi}{\Phi}^{A'R'}(V_\cF^{A'B' \rarr E'D'})^\sT \notag\\ 
            &\hspace{3pc}\ketbra{\Phi}{\Phi}^{DD'}(V_\cF^{A'B' \rarr E'D'})^* \notag\\
            &\hspace{4pc}\ketbra{\Phi}{\Phi}^{A'R'}(V_\cF^{A'B' \rarr E'D'})^\sT \\
            &=\f{d_Bd_E}{d_Ad_D}(V_\cF^{A'B' \rarr E'D'})^*\ket{\Phi}^{A'R'}\bra{\Phi}^{EE'}V_\cF^{AB\rarr ED} \notag \\
            &\hspace{3pc}\ketbra{\Phi}{\Phi}^{BB'}(V_\cF^{AB \rarr ED})^\dag \notag\\
            &\hspace{4pc}\ket{\Phi}^{EE'}\bra{\Phi}^{A'R'}(V_\cF^{A'B' \rarr E'D'})^\sT \\
            &=\f{d_B}{d_D}(V_\cF^{A'B' \rarr E'D'})^* \notag\\
            &\hspace{3pc}( \omega^{DB'} \otimes \Phi^{A'R'})(V_\cF^{A'B' \rarr E'D'})^\sT \\
            &=\f{d_B}{d_D}\omega_0^{DD'E'R'},
    \end{align}
where we used Lemma \ref{lem: rotation lemma} in the second equation.
We then compute $(\Lambda \Lambda^\dag \Lambda)^{DD'E'R'}$:
\begin{align}
    &(\Lambda\Lambda^\dag\Lambda)^{DD'E'R'} \notag\\
    &= (\Pi_2\Pi_1\Pi_2\Pi_1)^{DD'E'R'} \\
    &= \f{d_B}{d_D} \ketbra{\Phi}{\Phi}^{DD'} \omega_0^{DD'E'R'} \\ 
    \label{inteq:9}
    &=  \sum_{\mu=1}^r \f{d_B\lambda_\mu}{d_D} \ketbra{\Phi}{\Phi}^{DD'}\ketbra{\psi_\mu}{\psi_\mu}^{DD'E'R'}.
\end{align}
To proceed with the calculation, we must determine $\bra{\Phi}^{DD'}\ket{\psi_\mu}^{DD'E'R'}$, which is an unnormalized vector on $E'R'$.
Although this is somewhat involved, it can be obtained as
\begin{equation}
\label{eq: inner pro. eta}
    \bra{\Phi}^{DD'}\ket{\psi_\mu}^{DD'E'R'} = \sqrt{\f{d_B \lambda_\mu}{d_D}}\ket{\eta_\mu^*}^{E'R'}.
\end{equation}
by investigating the state after the post-selection as follows.

Let us denote by $\ket{\zeta_{\rm succ}}^{REE'R'}$ the state on $REE'R'$ after the post-selection by $\ket{\Phi}^{DD'}$, i.e., 
\begin{equation}
\label{eq: post state succ def}
    \ket{\zeta_{\rm succ}}^{REE'R'} = \f{1}{\sqrt{p_{\rm succ}}}\bra{\Phi}^{DD'}\ket{\omega_0}^{REDD'E'R'}.
\end{equation}
The states $\omega_0^{RE}$ and $\zeta_{\rm succ}^{RE}$ are related by 
\begin{align}
\label{eq: relation of zeta and omega0}
    \zeta_{\rm succ}^{RE} = \f{d_B}{d_D p_{\rm succ}} \big(\omega_0^{RE}\big)^2.
\end{align}
This follows from the direct calculation: 
\begin{align}
    &\zeta_{\rm succ}^{RE} \notag\\
    &= \f{1}{p_{\rm succ}} \bra{\Phi}^{DD'}\omega_0^{REDD'}\ket{\Phi}^{DD'} \\
    &= \f{1}{p_{\rm succ}} \tr_{R'E'}\Big[\big(\bra{\Phi}^{DD'}(V_\cF^{A'B'\rarr E'D'})^* \ket{\Phi}^{A'R'}\big)  \notag\\
    &\hspace{1pc}\ketbra{\omega}{\omega}^{REDB'}\big(\bra{\Phi}^{DD'}(V_\cF^{A'B'\rarr E'D'})^*\ket{\Phi}^{A'R'}\big)^\dag\Big] \\
    &= \f{d_B d_E}{d_A d_D p_{\rm succ}} \tr_{R'E'}\Big[\big(\bra{\Phi}^{\hat{B}B'}(V_\cF^{R'\hat{B}\rarr \hat{E}D})^\dag \notag\\
    &\hspace{6pc} \ket{\Phi}^{\hat{E}E'}\big)\ketbra{\omega}{\omega}^{REDB'} \notag\\
    &\hspace{4pc}\big(\bra{\Phi}^{\hat{B}B'}(V_\cF^{R'\hat{B}\rarr \hat{E}D})^\dag\ket{\Phi}^{\hat{E}E'}\big)^\dag\Big] \\
    &= \f{d_B}{d_A d_D p_{\rm succ}} \tr_{R'}\Big[\bra{\Phi}^{\hat{B}B'}(V_\cF^{R'\hat{B}\rarr \hat{E}D})^\dag \notag\\
    &\hspace{2pc}(\ketbra{\omega}{\omega}^{REDB'} \otimes \bI^{\hat{E}})(V_\cF^{R'\hat{B}\rarr \hat{E}D}) \ket{\Phi}^{\hat{B}B'}\Big] \\
    &= \f{d_B}{d_D p_{\rm succ}}\tr_{\hat{E}DB'}\Big[(V_\cF^{R'\hat{B}\rarr \hat{E}D}) (\pi^{R'} \otimes \Phi^{\hat{B}B'}) \notag\\
    &\hspace{2pc}(V_\cF^{R'\hat{B}\rarr \hat{E}D})^\dag (\ketbra{\omega}{\omega}^{REDB'} \otimes \bI^{\hat{E}})\Big] \\
    &= \f{d_B}{d_D p_{\rm succ}}\tr_{DB'}[\omega^{DB'} \ketbra{\omega}{\omega}^{REDB'}] \\
    &= \f{d_B}{d_D p_{\rm succ}}(\omega^{RE})^2 \\
    &= \f{d_B}{d_D p_{\rm succ}} (\omega_0^{RE})^2,
\end{align}
where we used Lemma~\ref{lem: rotation lemma} in the third equation.
The second last equality follows, e.g., from the Schmidt decomposition of $\ket{\omega}^{REDB'}$.
Note that $\omega^{RE} = \omega_0^{RE}$.

We now expand Eq.~\eqref{eq: relation of zeta and omega0} based on the Schmidt decomposition of $\ket{\omega_0}^{REDD'E'R'}$ given by Eq.~\eqref{eq: schmidt omega 0}. As $\ket{\zeta_{\rm succ}}^{REE'R'}$ is given by Eq.~\eqref{eq: post state succ def}, it can be written as
\begin{equation}
\label{eq: zeta unnormalize}
    \ket{\zeta_{\rm succ}}^{REE'R'} = \sum_{\mu=1}^r \sqrt{\f{\lambda_\mu}{p_{\rm succ}}}\ket{\eta_\mu}^{RE}\ket{\check{\varphi}_\mu}^{E'R'},
\end{equation}
where $\ket{\check{\varphi}_\mu}^{E'R'} = \bra{\Phi}^{DD'}\ket{\psi_\mu}^{DD'E'R'}$ is an unnormalized vector. Hence, we can rephrase Eq.~\eqref{eq: relation of zeta and omega0} as
\begin{multline}
\label{inteq:10}
    \sum_{\mu, \nu=1}^r \f{\sqrt{\lambda_\mu \lambda_\nu}}{p_{\rm succ}}\braket{\check{\varphi}_\nu}{\check{\varphi}_\mu} \ketbra{\eta_\mu}{\eta_\nu}^{RE} \\
    =
    \sum_{\mu=1}^r \f{d_B\lambda_\mu^2}{d_D p_{\rm succ}}\ketbra{\eta_\mu}{\eta_\mu}^{RE}.
\end{multline}
Since $\{\ket{\eta_\mu}^{RE}\}_\mu$ is an orthonormal basis and $\lambda_\mu \neq 0$ for $\mu = 1, 2, \ldots, r$, Eq.~\eqref{inteq:10} implies that
\begin{equation}
\label{Eq:sssssssss}
    \braket{\check{\varphi}_\nu}{\check{\varphi}_\mu}
    =
    \f{d_B\lambda_\mu}{d_D}\delta_{\mu \nu},
\end{equation}
where $\delta_{\mu \nu} = 1$ if $\mu=\nu$ and $\delta_{\mu \nu} =0$ otherwise.
By normalizing $\ket{\check{\varphi}_\mu}$, such as $\ket{\varphi_\mu}^{E'R'} = \sqrt{d_D/(d_B\lambda_\mu)}\ket{\check{\varphi}_\mu}^{E'R'}$, we obtain
\begin{equation}
    \begin{split}
    \label{inteq:11}
    \ket{\zeta_{\rm succ}}^{REE'R'} = \sum_{\mu=1}^r \sqrt{\f{d_B}{d_D p_{\rm succ}}}\lambda_\mu \ket{\eta_\mu}^{RE}\ket{\varphi_\mu}^{E'R'}.
    \end{split}
\end{equation}
Since $\ket{\varphi_\mu}^{E'R'}$ are mutually orthonormal due to Eq.~\eqref{Eq:sssssssss}, Eq.~\eqref{inteq:11} provides the Schmidt decomposition of $\ket{\zeta_{\rm succ}}^{REE'R'}$ between $RE$ and $E'R'$.

\begin{figure}
    \centering
    \includegraphics[width = 73mm]{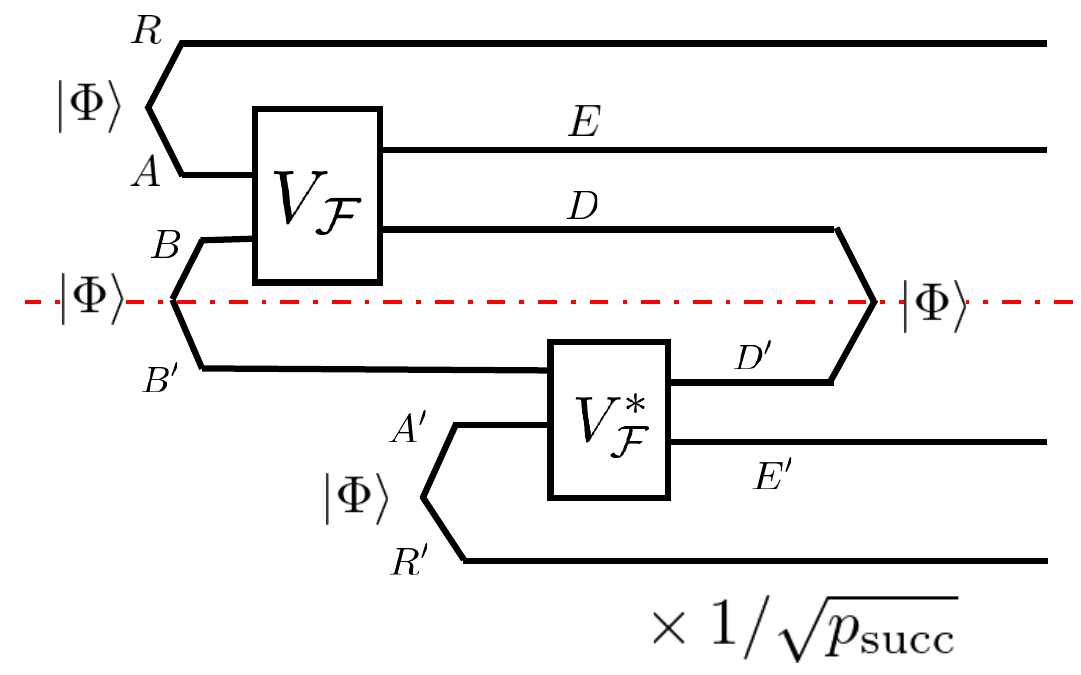}
    \caption{A diagram of the state $\ket{\zeta_{\rm succ}}^{REE'R'}$. This is symmetrical with respect to the red dash-dotted line, up to the complex conjugate.
    Due to this symmetry, the Schmidt basis of $\ket{\zeta_{\rm succ}}^{REE'R'}$ is given by $\{\ket{\eta_\mu}^{RE}\ket{\eta_\mu^*}^{E'R'}\}_\mu$.}
    \label{fig: symmetry}
\end{figure}

Moreover, from the symmetry of $\ket{\zeta_{\rm succ}}^{REE'R'}$ between $RE$ and $E'R'$, taking the complex conjugate of this state
is equivalent to swapping $RE$ with $E'R'$.
This implies that the Schmidt basis of $\ket{\zeta_{\rm succ}}^{REE'R'}$ in $RE$ and that in $E'R'$ are the same up to the complex conjugate.
Hence, we find that $\ket{\varphi_\mu}^{E'R'} = \ket{\eta_\mu^*}^{E'R'}$.
Accordingly, the Schmidt decomposition of $\ket{\zeta_{\rm succ}}^{REE'R'}$ is given by
\begin{align}
\label{eq: schmidt zeta true}
    \ket{\zeta_{\rm succ}}^{REE'R'} = \sum_{\mu=1}^r \sqrt{\f{d_B}{d_D p_{\rm succ}}} \lambda_\mu \ket{\eta_\mu}^{RE}\ket{\eta_\mu^*}^{E'R'}.
\end{align}
This symmetry property solely relies on the general structure of the protocol and is of crucial importance. See also Fig.~\ref{fig: symmetry}.

Finally, by applying $\bra{\eta_\mu}^{RE}$ on the left to $\ket{\zeta_{\rm succ}}^{REE'R'}$ in its two different expressions given by Eqs.~\eqref{eq: zeta unnormalize} and~\eqref{eq: schmidt zeta true}, we obtain Eq.~\eqref{eq: inner pro. eta}.\\

We now return to determining the singular value decomposition of $\Lambda^{DD'E'R'}$. Substituting Eq.~\eqref{eq: inner pro. eta} into Eq.~\eqref{inteq:9}, it follows that
\begin{align}
     &(\Lambda\Lambda^\dag\Lambda)^{DD'E'R'} \notag\\
     &= \sum_{\mu=1}^r \Big(\f{d_B\lambda_\mu}{d_D}\Big)^{3/2} \ket{\Phi}^{DD'}\ket{\eta_\mu^*}^{E'R'}\bra{\psi_\mu}^{DD'E'R'}.
\end{align}
Hence, the singular value decomposition of $\Lambda^{DD'E'R'}$ is uniquely determined, without any phase ambiguity in the left- and right-singular vectors, as
\begin{align}
\label{eq: singular value decomp Lambda}
    &\Lambda^{DD'E'R'} \notag\\
    &= \sum_{\mu=1}^r \sqrt{\f{d_B\lambda_\mu}{d_D}} \ket{\Phi}^{DD'}\ket{\eta_\mu^*}^{E'R'}\bra{\psi_\mu}^{DD'E'R'}.
\end{align}
When some of the $\lambda_\mu$'s are degenerate, we fix the basis in the degenerate subspace to an arbitrary one.

Since $\Lambda^{DD'E'R'}$ is a product of two projectors $\Pi_1^{D'E'R'}$ and $\Pi_2^{DD'}$, it is naturally block-encoded in $\bI^{DD'E'R'}$ on the subspace specified by these projectors; see also Eqs.~\eqref{inteq:5} to~\eqref{inteq:7} in~\ref{sec: explain QSVT FPAA}, with $U = \bI$.
Then, the following holds as an immediate consequence of the QSVT for $\Lambda^{DD'E'R'}$ with a real odd polynomial~\cite{gilyen2019qsvt, Gilyn2019thesis, martyn2021grand}.

Let $Q_{t, \phi}(x)$ be a degree-$t$ real odd polynomial satisfying $|Q_{t, \phi}(x)| \leq 1$ for all $x \in [-1, 1]$. Then, there is $\phi \in (-\pi, \pi]^t$ such that
\begin{align}
\label{eq: G_FPAA QSVT}
        &(\Pi_2^{DD'}\otimes \bra{0}^H)G_{t, \phi}^{DD'E'R'H}(\Pi_1^{D'E'R'}\otimes\ket{0}^H) \notag\\
        &= Q_{t, \phi}(\Lambda^{DD'E'R'}) \\
        &= \sum_{\mu=1}^r Q_{t, \phi}\Big(\sqrt{\f{d_B\lambda_\mu}{d_D}}\Big) \ket{\Phi}^{DD'}\ket{\eta_\mu^*}^{E'R'}\bra{\psi_\mu}^{DD'E'R'},
    \end{align}
where the unitary $G_{t, \phi}^{DD'E'R'H}$ is given by Eq.~\eqref{eq: FPAA unitary}, with a single-qubit auxiliary system $H$.
If we can choose the polynomial $Q_{t, \phi}$ such that $Q_{t, \phi}\big(\sqrt{d_B\lambda_\mu/d_D}\big) \approx 1$ for all $\mu = 1, 2, \ldots, r$, we obtain desired transformation $\ket{\psi_\mu}^{DD'E'R'}$ to $\ket{\Phi}^{DD'}\ket{\eta_\mu^*}^{E'R'}$. A possible choice of such a polynomial is one that approximates the sign function.
From Lemma~\ref{lem: poly apprx sign func}, for $Q_t^{\rm sign}\big(\sqrt{d_B\lambda_\mu/d_D}\big)$ to be larger than $1-\delta$ for $\mu = 1, 2, \ldots, r$, it suffices that $\sqrt{d_B\lambda_{\rm min}/d_D} \geq \beta$, where $\lambda_{\rm min} \coloneqq \min_{\mu \in [1, r]}\lambda_\mu$ and $\delta \in (0, 1/2)$.
Thus, we can take the odd integer $t$ such that  
\begin{align}
\label{eq: gyk t cond 2}
     t \geq \bigg\lceil8e\sqrt{\f{d_D}{d_B\lambda_{\rm min}(\omega^{RE})}}\log(2/\delta)\bigg\rceil.
\end{align}
We here used that $\lambda_{\rm min}$ corresponds to the non-zero minimum eigenvalue of $\omega^{RE}$, $\lambda_{\rm min}(\omega^{RE})$, since $\omega_0^{RE} = \omega^{RE}$; see Eq.~\eqref{eq: schmidt omega 0}.

This QSVT allows us to approximately obtain a block-encoding of the operator 
\begin{align}
&\sign(\Lambda^{DD'E'R'}) \otimes \ketbra{0}{0}^H 
\notag\\
&\hspace{0pc}=  \sum_{\mu=1}^r \ket{\Phi}^{DD'}\ket{\eta_\mu^*}^{E'R'}\bra{\psi_\mu}^{DD'E'R'} \otimes \ketbra{0}{0}^H.
\end{align}
Applying this to the state $\ket{\omega_0}^{REDD'E'R'}\ket{0}^H$, we achieve the transformation:
\begin{equation}
\label{eq: trans from omega0 to targ}
    \ket{\omega_0}^{REDD'E'R'}\ket{0}^H \mapsto \ket{\Phi}^{DD'}\ket{\omega_{\rm targ}}^{REE'R'}\ket{0}^H.
\end{equation}
Recall that the Schmidt decomposition of $\ket{\omega_0}^{REDD'E'R'}$ and $\ket{\omega_{\rm targ}}^{REE'R'}$ are given in Eqs.~\eqref{eq: schmidt omega 0} and~\eqref{eq: def omega targ}, respectively.
Since the state $\ket{\omega_{\rm targ}}^{REE'R'}$ satisfies Eq.~\eqref{eq: target state of lemma}, we can successfully recover the MES $\ket{\Phi}^{RR'}$ for sufficiently small $\epsilon$.\\

To provide a more precise analysis of the overall error, we denote the output state of the QSVT-based FPAA algorithm by 
\begin{equation}\begin{split}
    &\ket{\omega_t}^{REDD'E'R'H} \notag\\
    &\hspace{1pc}\coloneqq G_{t, \phi}^{DD'E'R'H}\ket{\omega_0}^{REDD'E'R'}\ket{0}^H.
\end{split}\end{equation}
By taking $t$ and $\phi$ to approximate the sign function, we obtain the overlap between this output state and the state $\ket{\Phi}^{DD'}\ket{\omega_{\rm targ}}^{REE'R'}\ket{0}^H$ as
\begin{align}
    &\bra{\Phi}^{DD'}\bra{\omega_{\rm targ}}^{REE'R'}\bra{0}^H\ket{\omega_t}^{REDD'E'R'H} \notag\\
    &=\sum_{\mu=1}^r \lambda_\mu \bra{\Phi}^{DD'}\bra{\eta_\mu^*}^{E'R'}\bra{0}^H  \notag\\
    &\hspace{4pc} G_{t, \phi}^{DD'E'R'H} \ket{\psi_\mu}^{DD'E'R'}\ket{0}^H \\
    &=\sum_{\mu=1}^r \lambda_\mu Q_t^{\rm sign}\big(\sqrt{d_B\lambda_\mu/d_D}\big) \\
    &\geq (1-\delta)\sum_{\mu=1}^r \lambda_\mu \\
    &= 1-\delta,
\end{align}
where we used $\sum_{\mu=1}^r \lambda_\mu =1$.
Using the Fuchs--van de Graaf inequalities and the contraction property of the trace norm under the partial trace, we obtain
\begin{equation}
\label{eq: ineq after qsvt}
    \frac{1}{2}\|\omega_t^{RR'} - \omega_{\rm targ}^{RR'} \|_1 
    \leq \sqrt{1-(1-\delta)^2} 
    \leq \sqrt{2 \delta}.
\end{equation}
Using Eqs.~\eqref{eq: target state of lemma},~\eqref{eq: ineq after qsvt}, and the triangle inequality, we obtain 
\begin{equation}
    \f{1}{2}\|\omega_t^{RR'} - \Phi^{RR'}\|_1 \leq \sqrt{\epsilon} + \sqrt{2 \delta},
\end{equation}
where $\epsilon$ is the degree of decoupling.
Note that the state $\omega^{RR'}_t$ is the output of the generalized YK decoder: $\cD_{t, \phi}^{DB'\rarr R'}(\omega^{RDB'}) = \omega^{RR'}_t$. Thus, by rescaling $\delta$ to $\delta^2/2$, we complete the evaluation of the overall recovery error of the generalized YK decoder.


We next investigate the circuit complexity of the generalized YK decoder. The non-trivial part is to implement the unitary $G_{t, \phi}$ in the QSVT-based FPAA algorithm, whose construction is explained in~\ref{sec: det GYK} (see also Fig.~\ref{fig: det GYK}). 
We thus mainly focus on $\cC(G_{t, \phi})$.

\begin{figure}
    \centering
    \includegraphics[width=80mm]{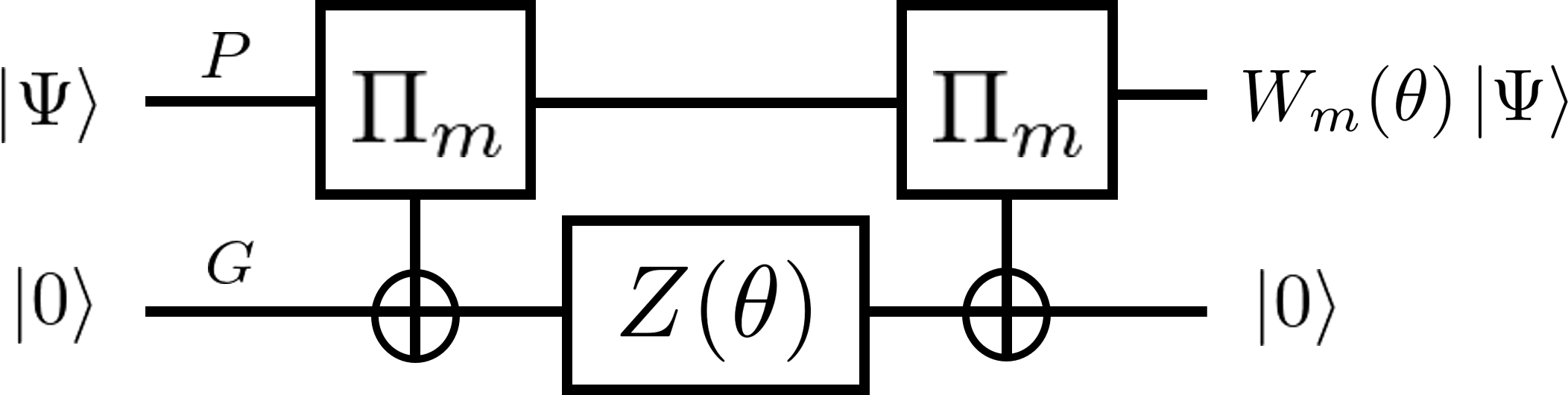}
    \caption{A quantum circuit for implementing a unitary $W_m(\theta)^P$. The box in which a projector is written implies that this projector controls the gate. The circle drawn inside the intersecting lines represents the NOT gate, i.e., the Pauli-$X$ gate.}
    \label{fig: FPAA1 circuit}
\end{figure}

We start with a circuit implementation of $W_m(\theta)$ for $m = 1, 2$:
\begin{align}
    W_m(\theta) 
    &= e^{i\theta(2\Pi_m -\mathbb{I})} \\
    &= e^{-i\theta}\mathbb{I} - (e^{-i\theta} - e^{i\theta})\Pi_m.
\end{align}
To implement the unitary $W_m(\theta)$, we use the \emph{projector-controlled NOT gate} \cite{gilyen2019qsvt, Gilyn2019thesis} that is in general defined for a projector $\Pi^P$ as
\begin{equation}\begin{split}
    {\rm C}_{\Pi}{\rm NOT}^{P\mathchar`-G} &\coloneqq \Pi^P \otimes X^G  + (\mathbb{I}^P - \Pi^P) \otimes \mathbb{I}^{G}.
\end{split}\end{equation}
The order of the superscripts in the left-hand side indicates the controlling and controlled systems. The gate $X$ is the single-qubit Pauli-$X$ gate.
We also use a single-qubit rotation-$Z$ gate:
\begin{align}
    Z(\theta) &\coloneqq e^{-i \theta Z} \\
    &= e^{-i\theta}\ketbra{0}{0} + e^{i\theta}\ketbra{1}{1},
\end{align}
where $Z$ is the single-qubit Pauli-$Z$ gate.
It is straightforward to check that, for any state $\ket{\Psi}^P$,
\begin{align}
    &({\rm C}_{\Pi}{\rm NOT}^{P\mathchar`-G}Z(\theta)^G{\rm C}_{\Pi}{\rm NOT}^{P\mathchar`-G})( \ket{\Psi}^P \otimes \ket{0}^G)  \notag\\
    &\hspace{0pc}= \big[e^{-i\theta}\mathbb{I}^P - (e^{-i\theta} - e^{i\theta})\Pi^P)\ket{\Psi}^P \big] \otimes \ket{0}^G \\
    &\hspace{0pc}= W_m(\theta)^P \ket{\Psi}^P \otimes \ket{0}^G.
\end{align}
Hence, we can implement $W_m(\theta)^P$ by preparing a single-qubit system $G$ and by operating a quantum circuit in Fig.~\ref{fig: FPAA1 circuit}.

\begin{figure}
    \centering
    \includegraphics[width=80mm]{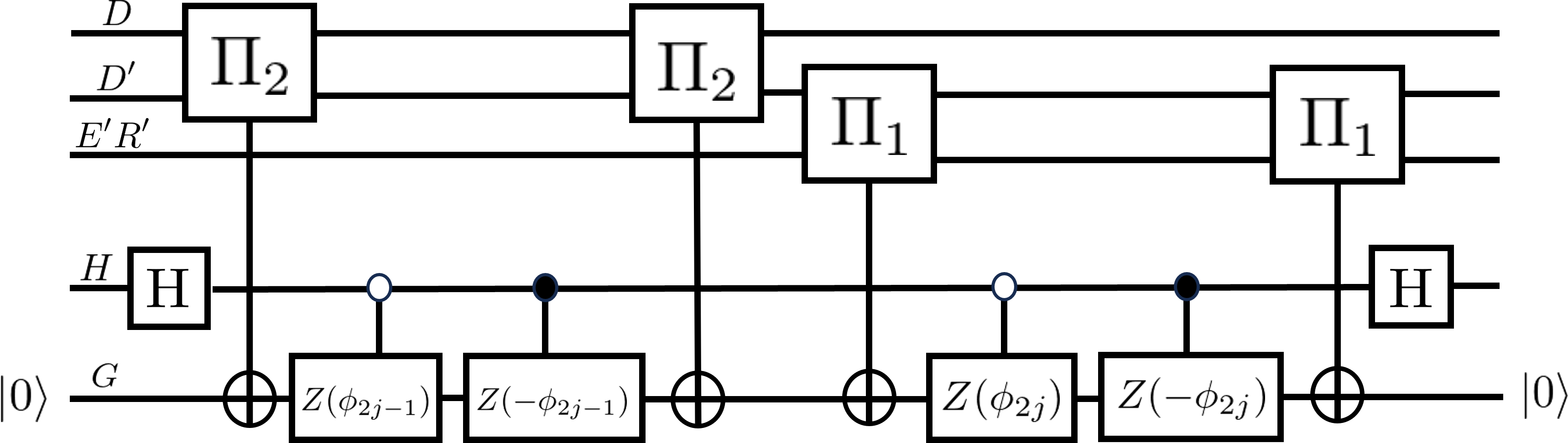}
    \caption{A quantum circuit for implementing the unitary operation
    $W_1(\phi_{2j})^{D'E'R'} W_2(\phi_{2j-1})^{DD'}  \otimes\ketbra{+}{+}^H + W_1(-\phi_{2j})^{D'E'R'}W_2(-\phi_{2j-1})^{DD'}\otimes\ketbra{-}{-}^H$. 
    Open circles imply that the gates are controlled by $\ket{0}$, while closed circles indicate the ones controlled by $\ket{1}$.}
    \label{fig: cell of FPAA}
\end{figure}

To construct a circuit for $G_{t, \phi}$, we prepare another single-qubit system $H$ for the controlled implementation of $W_m(\theta)^P$. For instance, a quantum circuit implementing 
\begin{equation}
    \begin{split}
        &W_1(\phi_{2j})^{D'E'R'}W_2(\phi_{2j-1})^{DD'}\otimes\ketbra{+}{+}^H  \\
        &\hspace{0.5pc}+ W_1(-\phi_{2j})^{D'E'R'}W_2(-\phi_{2j-1})^{DD'}\otimes\ketbra{-}{-}^H,
    \end{split}
\end{equation}
is given in Fig.~\ref{fig: cell of FPAA}. By applying the circuit $(t-1)/2$ times with various phases and finally applying $W_2(\phi_t)^{DD'} \otimes \mathrm{H}^H$, the unitary $G_{t, \phi}$ is realized. Here, the gate $\mathrm{H}^H$ is the single-qubit Hadamard gate on the system $H$.

\begin{figure}
    \centering
    \includegraphics[width=80mm]{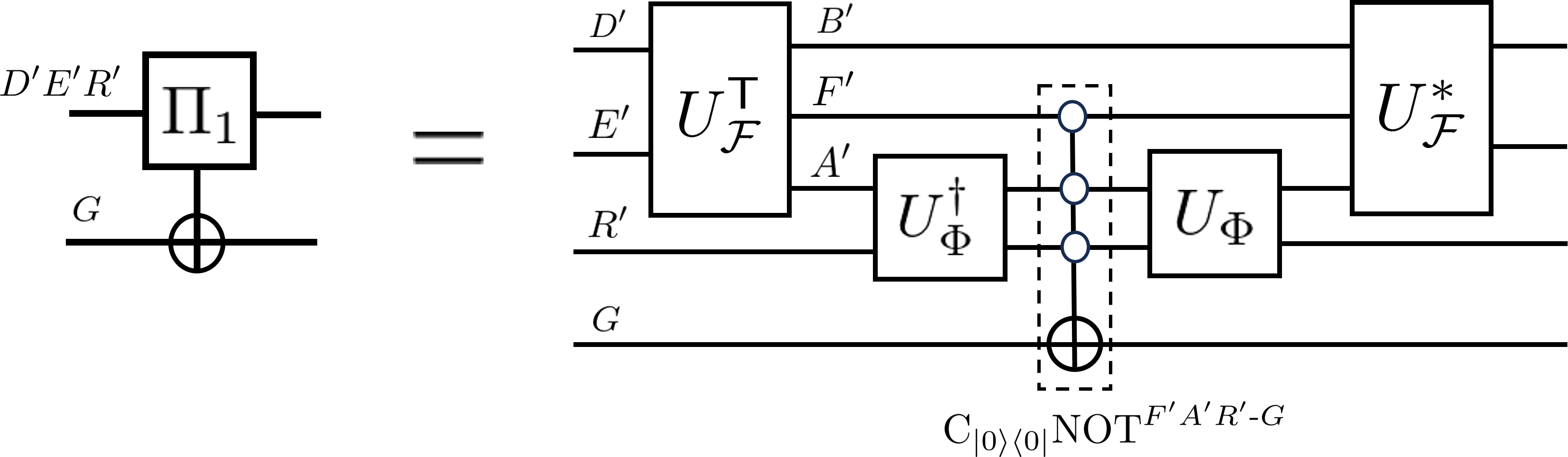}
    \caption{A quantum circuit for implementing the projector-controlled NOT gate ${\rm C}_{\Pi_1}{\rm NOT}^{D'E'R'\mathchar`-G}$. The dashed box represents the gate $\mathrm{C}_{\ket{0}\bra{0}}\mathrm{NOT}^{F'A'R'\mathchar`-G}$.}
    \label{fig: FPAA2 CpiNOT}
\end{figure}

In the construction, the unitary $G_{t, \phi}$ is decomposed into two unitaries $\mathrm{C}_{\Pi_1}\mathrm{NOT}^{D'E'R'\mathchar`-G}$ and $\mathrm{C}_{\Pi_2}\mathrm{NOT}^{DD'\mathchar`-G}$.
A quantum circuit for $\mathrm{C}_{\Pi_1}\mathrm{NOT}^{D'E'R'\mathchar`-G}$ is given in Fig.~\ref{fig: FPAA2 CpiNOT}. 
In general, the unitary $\mathrm{C}_{\ketbra{0}{0}}\mathrm{NOT}^{P\mathchar`-G}$ can be implemented using $\mathcal{O}(\log d_P)$ single- and two-qubit gates and $\mathcal{O}(\log d_P)$ ancilla qubits \cite{nielsen2010quantum}. The unitary $U_\Phi^{A'R'}$, which is given by
\begin{equation}
    U_{\Phi}^{A'R'}\ket{0}^{A'}\ket{0}^{R'} = \ket{\Phi}^{A'R'},
\end{equation}
can be implemented using $\cO(\log d_A)$ gates. 
Hence, in total, ${\rm C}_{\Pi_1}{\rm NOT}^{D'E'R'\mathchar`-G}$ can be implemented by 
\begin{equation}
   \cO\big(\cC(U_{\cF}) + \log{(d_Ad_F)}\big)
\end{equation}
gates and $\cO(\log d_Ad_F)$ ancilla qubits.~Similarly, ${\rm C}_{\Pi_{2}}{\rm NOT}^{DD'\mathchar`-G}$ can be implemented using $\cO(\log d_D)$ gates and $\cO(\log d_D)$ ancilla qubits. 

In the unitary $G_{t, \phi}$, these projector-controlled NOT gates are used $\cO(t)$ times. The total complexity of the generalized YK decoder is given by
\begin{align}
 \label{eq: comp gYK proof}
      \mathcal{C}(\mathcal{D}_{t, \phi})  &= \cO\Big(t  \big(\cC(U_{\cF}) + \log(d_Ad_Fd_D)\big)\Big) \notag\\
      &\hspace{3pc} + \cC(U_{\cF}) + \cO(\log d_A) \\
      &=\cO\Big(t\big(\cC(U_\cF) + \log(d_D^2d_E/d_B)\big)\Big),
\end{align}
with $\cO\big(\log{(d_D^2d_E/d_B)}\big)$ ancilla qubits.
Here, we used $d_Ad_Bd_F = d_Ed_D$.
In Eq.~\eqref{eq: comp gYK proof}, the first line in the right-hand side comes from $G_{t, \phi}$ and the second line comes from $\cV^{A'B' \rarr E'D'}$, which is applied before $G_{t, \phi}$.

\begin{figure*}
    \centering
    \includegraphics[width=160mm]{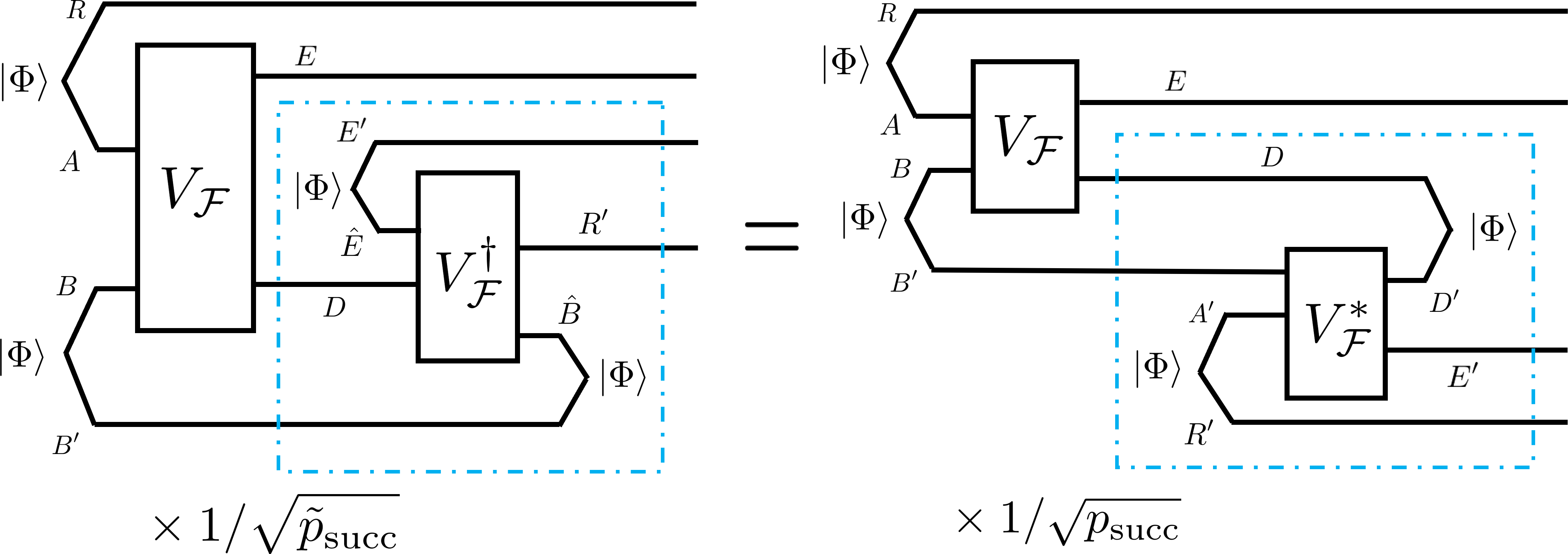}
    \caption{The equivalence of the states $\til{\zeta}_{\rm succ}^{REE'R'}$ and $\zeta_{\rm succ}^{REE'R'}$, which are obtained after the post-selection in the Petz-like protocol and in the generalized YK protocol, respectively. We can derive this equivalence by applying Lemma \ref{lem: rotation lemma} onto the portion enclosed by the blue dash-dotted lines.}
    \label{fig: IPMout=GYKout}
\end{figure*}

\subsection{Proofs: the Petz-like decoder}
\label{sec: proof of Plike}

Similarly to the generalized YK decoder, we first consider the decoding protocol with post-selection and then provide a sketch of a proof of Theorem \ref{thm: det IPM}.

From Eqs.~\eqref{eq: stinespring iso-uni},~\eqref{eq: like post map}, and~\eqref{eq: Plike post state}, the success probability $\til{p}_{\rm succ}$ is computed as
\begin{align}
    \til{p}_{\rm succ} &= \f{d_A}{d_E}\tr\big[V_\cF^{R'\hat{B} \rarr \hat{E}D}(\Phi^{\hat{B}B'} \otimes \pi^{R'}) \notag \\
    &\hspace{3pc}(V_\cF^{R'\hat{B} \rarr \hat{E}D})^\dag (\omega^{RDB'}\otimes \mathbb{I}^{\hat{E}})\big] \\
    &=\f{d_A}{d_E}\tr\big[(\omega^{DB'})^2\big] \\
    &=\f{d_A}{d_E}2^{-H_2(DB')_\omega} \\
    \label{eq: p rep. petzlike}
    &=\f{d_A}{d_E}2^{-H_2(RE)_\omega},
\end{align}
where we used $H_2(DB')_\omega = H_2(RE)_\omega$ as $\ket{\omega}^{REDB'}$ is pure. 
Note that the states $\omega^{RDB'}$ and $\ket{\omega}^{REDB'}$ are defined in Eqs.~\eqref{inteq:13} and~\eqref{inteq:12}, respectively.
The fidelity between $\til{\zeta}_{\rm succ}^{RR'}$ and $\Phi^{RR'}$ is computed as 
\begin{align}
        &\mathrm{F}(\til{\zeta}_{\rm succ}^{RR'}, \Phi^{RR'})  \notag \\
        &\hspace{1pc}= \f{1}{d_E \til{p}_{\rm succ}}\tr\big[V_\cF^{R'\hat{B}\rarr \hat{E}D}(\Phi^{RR'} \otimes\Phi^{\hat{B}B'})\notag \\
        &\hspace{5pc}(V_\cF^{R'\hat{B}\rarr \hat{E}D})^\dag(\omega^{RDB'}\otimes \mathbb{I}^{\hat{E}})\big] \\
        &\hspace{1pc}=\f{1}{d_A}2^{H_2(RE)_{\omega}}\tr\big[\omega^{R\hat{E}DB'}(\omega^{RB'D}\otimes \mathbb{I}^{\hat{E}})\big] \\
        &\hspace{1pc}=\f{1}{d_A}2^{H_2(RE)_\omega - H_2(RDB')_\omega} \\
        &\hspace{1pc}=\f{1}{d_A}2^{H_2(RE)_\omega - H_2(E)_\omega},
    \end{align}
by using $H_2(RDB')_\omega = H_2(E)_\omega$ for $\ket{\omega}^{REDB'}$.
Hence, we obtained Eqs.~\eqref{eq: post select P like} and~\eqref{eq: post fidelity plike}.


Let us now turn to the proof of Theorem~\ref{thm: det IPM}.
Since the proof is similar to that of Theorem~\ref{thm: the gYK}, we provide only an outline.

We denote the input state of the QSVT-based FPAA algorithm by
\begin{equation}
    \til{\omega}_0^{RE'R'\hat{F}\hat{B}B'} \coloneqq \til{\cV}^{D \rarr E'R'\hat{F}\hat{B}}(\omega^{RDB'}),
\end{equation}
where $\til{\cV}^{D \rarr E'R'\hat{F}\hat{B}}$ is the isometry map such that
\begin{equation}
    \til{\cV}^{D \rarr E'R'\hat{F}\hat{B}} = (U_\cF^{\hat{L}})^\dag(\cdot\otimes\Phi^{\hat{E}E'})U_\cF^{\hat{L}},
\end{equation}
and $\hat{L} = R'\hat{F}\hat{B} = \hat{E}D$.
Note that $\til{\omega}_0^{RE} = \omega^{RE}$.
Let $\ket{\til{\omega}_0}^{REE'R'\hat{F}\hat{B}B'}$ be the purified state such that
\begin{equation}
    \ket{\til{\omega}_0}^{REE'R'\hat{F}\hat{B}B'} = (U_\cF^{\hat{L}})^\dag\ket{\omega}^{REDB'}\ket{\Phi}^{\hat{E}E'}.
\end{equation}
The state on $REE'R'$ after the post-selection is then given by
\begin{align}
\label{eq: post select state plike}
    &\ket{\til{\zeta}_{\rm succ}}^{REE'R'} \notag\\
    &= \f{1}{\sqrt{\til{p}_{\rm succ}}}\bra{0}^{\hat{F}}\bra{\Phi}^{\hat{B}B'}\ket{\til{\omega}_0}^{REE'R'\hat{F}\hat{B}B'}.
\end{align}

It is important to note that 
\begin{equation}\begin{split}
\label{eq: equiv gYK and Plike post}
    \til{\zeta}_{\rm succ}^{REE'R'} = \zeta_{\rm succ}^{REE'R'},
\end{split}\end{equation}
where the right-hand side is the state after the post-selection in the generalized YK decoding protocol.
Although it may be hard to observe this relation from the construction in Fig.~\ref{fig: prob IPM}, this can be readily shown using Lemma \ref{lem: rotation lemma} as in Fig.~\ref{fig: IPMout=GYKout}. 
From this relation, it turns out that the state $\til{\zeta}_{\rm succ}^{REE'R'}$ is also symmetric between $RE$ and $E'R'$ up to the complex conjugate, and thus, the Schmidt basis in $RE$ and that in $E'R'$ are complex conjugates of each other.

Suppose that the Schmidt decomposition of $\ket{\til{\omega}_0}^{REE'R'\hat{F}\hat{B}B'}$, divided into $RE$ and $E'R'\hat{F}\hat{B}B'$, is given by 
\begin{equation}\begin{split}
\label{inteq:2}
     &\ket{\til{\omega}_0}^{REE'R'\hat{F}\hat{B}B'}  = \sum_{\mu=1}^{r}\sqrt{\lambda_\mu}\ket{\eta_\mu}^{RE}\ket{\tilde{\psi}_\mu}^{E'R'\hat{F}\hat{B}B'}.
\end{split}\end{equation}
As $\til{\omega}_0^{RE}$ is equal to $\omega_0^{RE} = \omega^{RE}$, $\lambda_\mu$ and $\ket{\eta_\mu}^{RE}$ are the eigenvalues and eigenvectors of $\omega^{RE}$, respectively.
Since the state $\ket{\til{\zeta}_{\rm succ}}$ is defined by using $\ket{\til{\omega}_0}$ as Eq.~\eqref{eq: post select state plike}, it follows that
\begin{equation}\begin{split}
    &\ket{\til{\zeta}_{\rm succ}}^{REE'R'}  \\
    &\hspace{1pc}= \sum_{\mu=1}^{r}\sqrt{\f{\lambda_\mu}{\til{p}_{\rm succ}}}\ket{\eta_\mu}^{RE}\bra{0}^{\hat{F}}\bra{\Phi}^{\hat{B}B'}\ket{\tilde{\psi}_\mu}^{E'R'\hat{F}\hat{B}B'}.
\end{split}\end{equation}
From Eq.~\eqref{eq: schmidt zeta true} for $\ket{\zeta_{\rm succ}}$ in the generalized YK decoding protocol with post-selection and the relation in Eq.~\eqref{eq: equiv gYK and Plike post}, we have
\begin{align}
    &\bra{0}^{\hat{F}}\bra{\Phi}^{\hat{B}B'}\ket{\tilde{\psi}_\mu}^{E'R'\hat{F}\hat{B}B'} \notag\\
    &\hspace{1pc}= \sqrt{\f{d_B \til{p}_{\rm succ} \lambda_\mu}{d_D p_{\rm succ}}}\ket{\eta_\mu^*}^{E'R'} \\
    \label{inteq:1}
    &\hspace{1pc}=\sqrt{\f{d_A \lambda_\mu}{d_E}}\ket{\eta_\mu^*}^{E'R'}.
\end{align}
Here, we substituted the success probabilities $p_{\rm succ}$ and $\tilde{p}_{\rm succ}$ in the generalized YK and Petz-like decoding protocols with post-selection, which are given by Eqs.~\eqref{eq: p rep.} and~\eqref{eq: p rep. petzlike}, respectively.

We next consider the QSVT-based FPAA.
The target state is, as in the case of the generalized YK decoder, given by $\ket{\omega_{\rm targ}}^{RER'E'}$ in Eq.~\eqref{eq: def omega targ}.
Recall that this state satisfies 
\begin{align}
    \label{inteq:3}
    \f{1}{2}\big\|\omega_{\rm targ}^{RR'} - \Phi^{RR'}\big\|_1 \leq \sqrt{\epsilon},
\end{align}
when the decoupling condition is satisfied with an error $\epsilon$.
Let $\til{\Pi}_1^{E'R'\hat{F}\hat{B}}$ and $\til{\Pi}_2^{\hat{F}\hat{B}B'}$ be projectors defined by
\begin{align}
\label{eq: two projection of Plike 2}
    &\til{\Pi}_1^{E'R'\hat{F}\hat{B}} 
    \coloneqq (U_\cF^{\hat{L}})^\dag (\ketbra{\Phi}{\Phi}^{\hat{E}E'}  \otimes \mathbb{I}^{D})U_\cF^{\hat{L}}, \\
\label{eq: prod 2 plike 2}
    &\til{\Pi}_2^{\hat{F}\hat{B}B'} \coloneqq \ketbra{0}{0}^{\hat{F}} \otimes \ketbra{\Phi}{\Phi}^{\hat{B}B'}.
\end{align}
From a similar calculation to the case of the generalized YK decoder, we obtain
\begin{align}
\label{eq: Pi product Plike}
    (\tilde{\Pi}_{1}\tilde{\Pi}_{2}\tilde{\Pi}_{1})^{E'R'\hat{F}\hat{B}B'} = \f{d_A}{d_E}\til{\omega}_0^{E'R'\hat{F}\hat{B}B'}.
\end{align}
Then, letting $\til{\Lambda}^{E'R'\hat{F}\hat{B}B'} = (\til{\Pi}_2\til{\Pi}_1)^{E'R'\hat{F}\hat{B}B'}$, it follows that
\begin{align}
    &\big(\til{\Lambda}\til{\Lambda}^\dag\til{\Lambda}\big)^{E'R'\hat{F}\hat{B}B'} \notag\\
    &= \ketbra{0}{0}^{\hat{F}} \otimes \ketbra{\Phi}{\Phi}^{\hat{B}B'} \f{d_A}{d_E}\omega_0^{E'R'\hat{F}\hat{B}B'} \\
    &= \sum_{\mu=1}^r \Big(\f{d_A \lambda_\mu}{d_E}\Big)^{3/2} \ket{\eta_\mu^*}^{E'R'} \notag\\
    &\hspace{6pc}\ket{0}^{\hat{F}}\ket{\Phi}^{\hat{B}B'}\bra{\til{\psi}_\mu}^{E'R'\hat{F}\hat{B}B'},
\end{align}
where we used Eqs.~\eqref{inteq:2},~\eqref{inteq:1}, and~\eqref{eq: Pi product Plike}.
Hence, the Schmidt decomposition of $\til{\Lambda}^{E'R'\hat{F}\hat{B}B'}$ is uniquely determined as
\begin{align}
    &\til{\Lambda}^{E'R'\hat{F}\hat{B}B'} \notag\\
    &= \sum_{\mu=1}^r \sqrt{\f{d_A \lambda_\mu}{d_E}} \ket{\eta_\mu^*}^{E'R'}\ket{0}^{\hat{F}}\ket{\Phi}^{\hat{B}B'}\bra{\til{\psi}_\mu}^{E'R'\hat{F}\hat{B}B'}.
\end{align}

By the QSVT-based FPAA algorithm with appropriately chosen $\phi \in (-\pi, \pi]^{t}$, we obtain a block-encoding $\til{G}_{t, \phi}^{E'R'\hat{F}\hat{B}B'H}$ of the operator
\begin{align}
    &\sign(\til{\Lambda}^{E'R'\hat{F}\hat{B}B'}) \otimes \ketbra{0}{0}^H 
    \notag\\
    &\hspace{0pc}=  \sum_{\mu=1}^r \ket{\eta_\mu^*}^{E'R'}\ket{0}^{\hat{F}}\ket{\Phi}^{\hat{B}B'} \bra{\til{\psi}_\mu}^{E'R'\hat{F}\hat{B}B'}\otimes \ketbra{0}{0}^H.
\end{align}
This operator maps the state $\ket{\til{\psi}_\mu}^{E'R'\hat{F}\hat{B}B'}\ket{0}^H$ to $\ket{\eta_\mu^*}^{E'R'}\ket{0}^{\hat{F}}\ket{\Phi}^{\hat{B}B'}\ket{0}^H$ for each $\mu$. 
Hence, the transformation:
\begin{align}
    &\ket{\til{\omega}_0}^{REE'R'\hat{F}\hat{B}B'}\ket{0}^H \notag\\
    &\hspace{2pc}\mapsto \ket{\omega_{\rm targ}}^{REE'R'}\ket{0}^{\hat{F}}\ket{\Phi}^{\hat{B}B'}\ket{0}^H,
\end{align}
is approximately achieved.
By a similar technique to the generalized YK decoder, we can conclude that the Petz-like decoder $\til{\cD}_{t, \phi}$ achieves
\begin{equation}
\label{eq: Plike after FPAA}
    \f{1}{2}\|\til{\cD}_{t, \phi}^{DB' \rarr R'}(\omega^{RDB'})-\omega_{\rm targ}^{RR'}\|\leq\sqrt{\delta},
\end{equation}
for $\delta \in (0, 1/2)$, where $t$ is any odd number satisfying
\begin{align}
        t \geq \bigg\lceil 8e\sqrt{\f{d_E}{d_A\lambda_{\rm min}(\omega^{RE})}}\log{(2/\delta)}\bigg\rceil.
\end{align}
We here used that $\lambda_{\rm min} = \lambda_{\rm min}(\omega^{RE})$.
From Eqs.~\eqref{inteq:3} and~\eqref{eq: Plike after FPAA}, by using the triangle inequality, the recovery error by the Petz-like decoder is evaluated as
\begin{equation}
    \f{1}{2}\|\til{\cD}_{t, \phi}^{DB'\rarr R'}(\omega^{RDB'}) - \Phi^{RR'}\|_1 \leq \sqrt{\epsilon} + \sqrt{2\delta}.
\end{equation}

Finally, since $\til{\Pi}_1^{E'R'\hat{F}\hat{B}}$ and $\til{\Pi}_2^{\hat{F}\hat{B}B'}$ are explicitly given by Eqs.~\eqref{eq: two projection of Plike 2} and~\eqref{eq: prod 2 plike 2}, respectively, the complexity of the Petz-like decoder can be evaluated similarly to the generalized YK decoder. The circuit complexity of $\mathrm{C}_{\til{\Pi}_1}\mathrm{ NOT}$ is 
\begin{equation}
    \cO\Big(\cC(U_\cF) + \log{d_E}\Big),
\end{equation}
and that of $\mathrm{C}_{\til{\Pi}_2}\mathrm{ NOT}$ is 
\begin{equation}
    \cO\big(\log{d_Bd_F}\big).
\end{equation}
Since they are applied $\cO(t)$ times in the Petz-like decoder, the total complexity is given by 
\begin{equation}
    \cO\Big(t\big(\cC(U_\cF) + \log{(d_Ed_Bd_F)}\big)\Big),
\end{equation}
with $\cO\big(\log{(d_Ed_Bd_F)}\big)$ ancilla qubits. 

By using $d_Ad_Bd_F = d_Ed_D$ and rescaling $\delta$ to $\delta^2/2$, Theorem~\ref{thm: det IPM} is obtained.

\section{Summary and outlooks}
\label{sec: conclusion}

In this paper, we have presented two decoders applicable to arbitrary noisy channels with explicit circuit implementations: the generalized YK decoder and the Petz-like decoder.
Both decoders can recover quantum information when the decoupling condition is satisfied, and they are applicable to entanglement-assisted and non-assisted settings.
They are thus among the first few explicit decoders that can be used to asymptotically achieve communication rates arbitrarily close to the quantum capacity or the entanglement-assisted quantum capacity, with suitably chosen encoders.

Both decoders are constructed in two steps: first we consider a decoding protocol with measurement and post-selection, and then replace the measurement with the QSVT-based FPAA algorithm to construct the decoder.
The two-step construction does not work with other AA-type algorithms.
Hence, our constructions of decoders provide an explicit example that fully leverages and highlights the unique strength of the QSVT-based FPAA in a practical problem.

We have also investigated their circuit complexity, showing that both decoders significantly reduce the computational cost compared to the previously known algorithmic implementation of the Petz recovery map. 
Moreover, we have shown that the generalized YK decoder has smaller complexity than the Petz-like decoder in many situations.

As a future direction, investigating protocols in the presence of circuit-level noise, i.e., noise that occurs during operations, is important. 
Since it is common, in studies of theoretical limits such as the quantum capacity, to assume that every operation except for the noisy channel can be realized noiselessly, we focused on the situation without the circuit-level noise. On the other hand, research that takes the circuit-level noise into account is also emerging~\cite{christ2024fault}. In our protocol, the QSVT-based FPAA may suffer from noisy operations, as mentioned in~\cite{Gilyn2019thesis}, since the effect of the noise may accumulate by iterating the operations.
To avoid the accumulation, it is desirable to iterate the operations as few times as
possible. Recalling that the number of iterations in the generalized YK decoder is related to the amount of the pre-shared entanglement (see Eq.~\eqref{eq: condition for t GYK}), the error accumulation is also related to the noise on the pre-share entanglement, making the investigation non-trivial. It is interesting to estimate the performance of the proposed decoders in the presence of the circuit-level noise.

From a theoretical viewpoint, it may also be interesting to address the question about whether a similar approach to that in this work functions for recovering \emph{classical}~\cite{schumacher1997sending, holevo1998capacityclassical} or \emph{hybrid}~\cite{Devetak2005capacityQchannel, hsieh2010entanglement, nakata2021oneshot, wakakuwa2023oneshot} information. In the former, the encoded information is classical, and the decoder is simply given by a quantum measurement. In the latter, the information is a mixture of classical and quantum, which can be decoded by the simultaneous use of a quantum measurement and a quantum decoder. In these cases, a couple of quantum measurements are known to work well, such as the pretty-good measurement~\cite{hausladen1994prettygood, holevo1998capacityclassical}. Our approach adapted to these settings may provide a better decoding performance.

Another direction is to relax the assumptions on the knowledge of the noisy channel~\cite{bjelakovic2009entanglement}.
While general decoders, including the proposed decoders, are constructed based on the assumption that we know the description of the noisy channel, it would not be realistic to obtain complete knowledge of the noise.
If we can relax the assumption, the decoders may become more useful.

These decoders may also have potential use in fundamental physics for exploring exotic quantum many-body phenomena that are related to the recovery of quantum information. For instance, the proposed decoders could be potentially applied to reconstructing the internal structure of a black hole from the noisy Hawking radiation \cite{Bao2020noisyhawking}, and to recovering the bulk structure from a part of the boundaries, such as the entanglement wedge reconstruction \cite{chen2020entanglement}. This is also an intriguing direction of studies with the decoders.

\acknowledgments

T. U. and Y. N. were supported by JST CREST Grant Number JPMJCR23I3.
T. U. was supported by JST SPRING Grant Number JPMJSP2108.
Y. N. was supported by MEXT-JSPS Grant-in-Aid for Transformative Research Areas (A) ”Extreme Universe” Grant Numbers JP21H05182 and JP21H05183, by JSPS KAKENHI Grant Number JP22K03464, and JST PRESTO Grant Number JPMJPR2456. The authors thank Takaya Matsuura, Shiro Tamiya, and Ryuji Takagi for valuable discussions.

\bibliographystyle{quantum}
\bibliography{ref}

\clearpage

\end{document}